\documentclass[aps, pra, reprint, superscriptaddress, floatfix]{revtex4-1} 

\usepackage[free-standing-units=true]{siunitx}
\usepackage{xspace} 
\usepackage{graphicx}
\usepackage{dsfont}
\usepackage{epstopdf}
\usepackage{bm, amsmath, amssymb}
\usepackage{float}
\usepackage{placeins}
\usepackage{xcolor}
\usepackage[T1]{fontenc}
\usepackage{boldline,multirow}

\newcommand{\sub}[1]{_{\mathrm{#1}}}
\newcommand{\ket}[1]{\left|#1\right>}
\newcommand{\bra}[1]{\left<#1\right|}

\newcommand{\fres}{\SI{8.9}{\giga\hertz}\xspace}
\newcommand{\fqubit}{\SI{6}{\giga\hertz}\xspace}
\newcommand{\kerr}{\SI{6.7}{\mega\hertz}\xspace}
\newcommand{\epstwofigtwo}{\SI{15.5}{\mega\hertz}\xspace}
\newcommand{\epstworest}{\SI{17.75}{\mega\hertz}\xspace}
\newcommand{\nbarfigtwo}{2.2\xspace}
\newcommand{\nbarRest}{2.6\xspace}
\newcommand{\bareTone}{\SI{15.5}{\micro\second}\xspace}

\newcommand{\bareTphi}{\SI{3.4}{\micro\second}\xspace}
\newcommand{\bareTecho}{\SI{13.7}{\micro\second}\xspace}
\newcommand{\catTphi}{\SI{105}{\micro\second}\xspace}
\newcommand{\catTphim}{\SI{106}{\micro\second}\xspace}

\newcommand{\tanhpump}{\SI{320}{\nano\second}\xspace}
\newcommand{\tanhrabi}{\SI{80}{\nano\second}\xspace}
\newcommand{\epsx}{\SI{740}{\kilo\hertz}\xspace}
\newcommand{\epsxGate}{\SI{6.5}{\mega\hertz}\xspace}
\newcommand{\xgateT}{\SI{24}{\nano\second}\xspace}

\newcommand{\zgateT}{\SI{38}{\nano\second}\xspace}
\newcommand{\qsT}{\SI{3.6}{\micro\second}\xspace}
\newcommand{\fid}{0.74\xspace}
\newcommand{\tauZp}{\SI{2.60}{\micro\second}\pm\SI{0.07}{\micro\second}\xspace}
\newcommand{\tauZm}{\SI{2.56}{\micro\second}\pm\SI{0.07}{\micro\second}\xspace}
\newcommand{\tauYp}{\SI{2.51}{\micro\second}\pm\SI{0.06}{\micro\second}\xspace}
\newcommand{\tauYm}{\SI{2.60}{\micro\second}\pm\SI{0.05}{\micro\second}\xspace}

\newcommand{\sdcoh}{\pm\SI{1}{\micro\second}\xspace}

\newcommand{\trise}{\SI{4}{\nano\second}\xspace}
\newcommand{\g}{\SI{1.7}{\mega\hertz}\xspace}
\newcommand{\kappatot}{\SI{1.9}{\mega\hertz}\xspace}
\newcommand{\kappaout}{\SI{1.4}{\mega\hertz}\xspace}
\newcommand{\kappaother}{\SI{0.5}{\mega\hertz}\xspace}
\newcommand{\kappapin}{\SI{80}{\kilo\hertz}\xspace}

\newcommand{\probmp}{0.13}
\newcommand{\probpm}{0.13}
\newcommand{\qndness}{0.85}
\newcommand{\epop}{\SI{4}{\percent}}
\newcommand{\detuning}{\SI{2.2}{\mega \hertz}}
\newcommand{\phiSNAIL}{0.26\,\Phi_0}
\newcommand{\gthree}{\SI{20}{\mega \hertz}}
\newcommand{\nth}{0.04}

\newcommand{\tauRB}{35.7 \xspace}
\newcommand{\rbError}{0.014 \xspace}
\newcommand{\rbFid}{98.6 \xspace} 

\newcommand{\calC}{\mathcal{C}}
\newcommand{\calN}{\mathcal{N}}
\newcommand{\xp}{\ket{+\alpha}}
\newcommand{\xm}{\ket{-\alpha}}

\newcommand{\ym}{\ket{\calC_{\alpha}^{+i}}}
\newcommand{\ypm}{\ket{\calC_{\alpha}^{\mp i}}}
\newcommand{\xpm}{\ket{\pm\alpha}}
\newcommand{\catp}{\ket{\calC_{\alpha}^+}}
\newcommand{\catm}{\ket{\calC_{\alpha}^-}}
\newcommand{\catpm}{\ket{\calC_{\alpha}^\pm}}
\newcommand{\catmp}{\ket{\calC_{\alpha}^\mp}}

\newcommand{\bloch}{a\xspace}
\newcommand{\blochfock}{b\xspace}
\newcommand{\phasespace}{c\xspace}
\newcommand{\cavity}{d\xspace}
\newcommand{\snailmon}{e\xspace}
\newcommand{\snail}{f\xspace}
\newcommand{\pulseRabi}{a\xspace}
\newcommand{\Rabivsamp}{b\xspace}
\newcommand{\Rabivsphase}{c\xspace}
\newcommand{\cutsRabivsphase}{d\xspace}
\newcommand{\wigners}{e\xspace}
\newcommand{\pulseMap}{a\xspace}
\newcommand{\pulseX}{c\xspace}
\newcommand{\pulseZ}{e\xspace}
\newcommand{\pulseQswitch}{b\xspace}
\newcommand{\blochMap}{b\xspace}
\newcommand{\blochX}{d\xspace}
\newcommand{\blochZ}{f\xspace}
\newcommand{\qswitch}{a\xspace}
\newcommand{\pulseTauZ}{f\xspace}
\newcommand{\pulseTauY}{d\xspace}
\newcommand{\tauZ}{g\xspace}
\newcommand{\tauY}{e\xspace}
\newcommand{\tauX}{c\xspace}

\newcommand{\SIFIGspec}{a\xspace}
\newcommand{\SIFIGpulsespec}{b\xspace}
\newcommand{\SIFIGcatPspec}{c\xspace}
\newcommand{\SIFIGcatMspec}{d\xspace}

\newcommand{\eqone}{1}
\newcommand{\eqtwo}{2}
\newcommand{\eqthree}{3}
\newcommand{\figone}{1}
\newcommand{\figtwo}{2}
\newcommand{\figthree}{3}
\newcommand{\figfour}{4}

\newcommand{\figwidth}{89mm} 
\newcommand{\figwidthWide}{183mm}

\newenvironment{ac}
{\begin{center}
		\textsc{author contributions}\\[1ex]
	\end{center}
}

\newenvironment{ack}
{\begin{center}
		\textsc{acknowledgements}\\[1ex]
	\end{center}
}

\begin{document}

\title{The Kerr-Cat Qubit: Stabilization, Readout, and Gates.}

\author{A. Grimm}
\thanks{These authors contributed equally.}
\email{alexander.grimm@psi.ch}
\affiliation{Department of Applied Physics, Yale University, New Haven, CT 06520, USA}
\author{N. E. Frattini}
\thanks{These authors contributed equally.}
\affiliation{Department of Applied Physics, Yale University, New Haven, CT 06520, USA}
\author{S. Puri}
\affiliation{Department of Physics, Yale University, New Haven, CT 06520, USA}
\author{S. O. Mundhada}
\affiliation{Department of Applied Physics, Yale University, New Haven, CT 06520, USA}
\author{S. Touzard}
\affiliation{Department of Applied Physics, Yale University, New Haven, CT 06520, USA}
\author{M. Mirrahimi}
\affiliation{QUANTIC team, Inria Paris, 2 rue Simone Iff, 75012 Paris, France}
\author{S. M. Girvin}
\affiliation{Department of Physics, Yale University, New Haven, CT 06520, USA}
\author{S. Shankar}
\thanks{Present address: Electrical and Computer Engineering, University of Texas, Austin}
\affiliation{Department of Applied Physics, Yale University, New Haven, CT 06520, USA}
\author{M. H. Devoret}
\email{michel.devoret@yale.edu}
\affiliation{Department of Applied Physics, Yale University, New Haven, CT 06520, USA}

\date{\today}

\begin{abstract}
Quantum superpositions of macroscopically distinct classical states, so-called Schr\"{o}dinger cat states, are a resource for quantum metrology, quantum communication, and quantum computation.
In particular, the superpositions of two opposite-phase coherent states in an oscillator encode a qubit protected against phase-flip errors~\cite{Cochrane1999,Mirrahimi2014}.
However, several challenges have to be overcome in order for this concept to become a practical way to encode and manipulate error-protected quantum information. The protection must be maintained by stabilizing these highly excited states and, at the same time, the system has to be compatible with fast gates on the encoded qubit and a quantum non-demolition readout of the encoded information. Here, we experimentally demonstrate a novel method for the generation and stabilization of Schr\"{o}dinger cat states based on the interplay between Kerr nonlinearity and single-mode squeezing~\cite{Cochrane1999,Milburn1991} in a superconducting microwave resonator~\cite{Puri2017}.
We show an increase in transverse relaxation time of the stabilized, error-protected qubit over the single-photon Fock-state encoding by more than one order of magnitude. We perform all single-qubit gate operations on time-scales more than sixty times faster than the shortest coherence time and demonstrate single-shot readout of the protected qubit under stabilization. Our results showcase the combination of fast quantum control with the robustness against errors intrinsic to stabilized macroscopic states and open up the possibility of using these states as resources in quantum information processing~\cite{Puri2018a,Puri2019,Guillaud2019,Goto2016b}.
\end{abstract}
\maketitle

A quantum system that can be manipulated and measured tends to interact with uncontrolled degrees of freedom in its environment leading to decoherence. This presents a challenge to the experimental investigation of quantum effects and in particular to the field of quantum computing, where quantum bits (qubits) must remain coherent while operations are performed. Most noisy environments are only locally correlated and thus cannot decohere quantum information encoded in a non-local manner. Therefore, some approaches to protect quantum information employ non-locality through the use of entangled qubit states~\cite{Shor1995}, spatial distance~\cite{Kitaev2003,Oreg2010,Lutchyn2010}, or a combination of both~\cite{Fowler2012a}. Crucially, this concept can be extended to states that are non-local in the phase space of a single oscillator~\cite{Gottesman2001a, Mirrahimi2014}, with the additional benefit of involving fewer physical components, a property that has been termed hardware efficiency. The latter is particularly relevant because fully protecting a quantum system against all forms of decoherence is likely to involve several layers of encodings, and it is crucial to introduce efficient error-protection into the physical layer while maintaining simplicity~\cite{Vuillot2019, Guillaud2019, Puri2019}.

\begin{figure}
 \includegraphics[angle = 0, width = \figwidth]{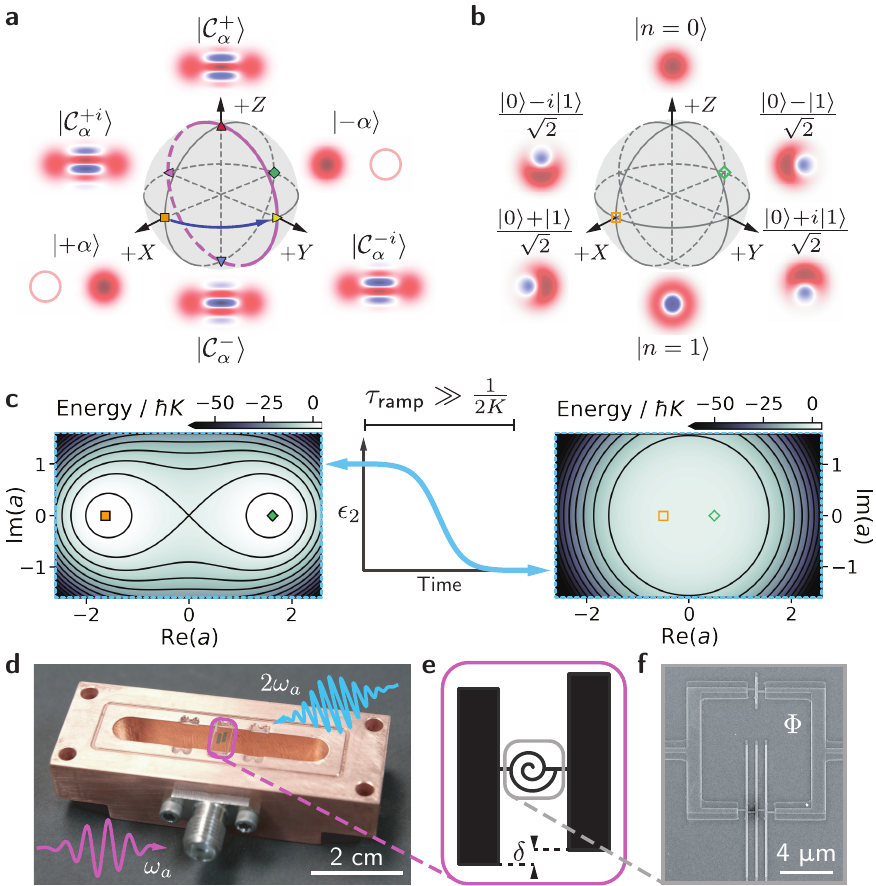}
  \caption{\label{fig1} \textbf{Qubit encoding, stabilization and implementation. a}, Bloch sphere of the protected ``Kerr-cat qubit'' in the large-$\alpha$ limit~\cite{SI}. The states on all six cardinal points are labelled, indicated by markers and their Wigner function~\cite{Haroche2006} phase-space representations are sketched. Here, $\ket{\pm Z}={|\mathcal{C}_{\alpha}^{\pm}\rangle = (|+\alpha\rangle \pm |-\alpha\rangle)/\sqrt{2}}$ and $\ket{\pm Y}={|\mathcal{C}_{\alpha}^{\mp i}\rangle =  (|+\alpha\rangle \mp i |-\alpha\rangle)/\sqrt{2}}$. The continuous $X(\theta)$ gate and the discrete $Z(\pi/2)$ gate are shown by a purple circle and a blue arrow, respectively. \textbf{b}, Bloch sphere of the single-photon ``Fock qubit'' shown for comparison with \textbf{a}. \textbf{c}, Energy dependence of equation~(\ref{eq:H}) on classical phase-space coordinates $\mathrm{Re}(a)$ and $\mathrm{Im}(a)$ for squeezing drive amplitudes $\epsilon_2/2\pi=\epstworest$ (left) and $\epsilon_2=0$ (right) with a sketch showing the adiabatic ramp of the drive over a time $\tau\sub{ramp}\gg1/2K$. Black lines are constant energy contours. The quadrature expectation values of the $|\pm X\rangle$-states from \textbf{a}, \textbf{b} are indicated by their respective markers. \textbf{d}, Photograph of the nonlinear resonator inside the copper section of the readout cavity~\cite{SI}. Also represented are the qubit-rotation drive ($\omega_a$) and the squeezing-generation drive ($2\omega_a$). \textbf{e}, Schematic of the nonlinear resonator. Large pads provide the capacitance and a superconducting nonlinear asymmetric inductive element (SNAIL)~\cite{Frattini2017} provides the nonlinear inductance. The pad offset $\delta$ sets the dispersive coupling to the readout mode. \textbf{f}, Scanning electron micrograph of the SNAIL (spiral symbol in panel \textbf{e}), consisting of four Josephson junctions in a loop threaded by an external magnetic flux $\Phi$.}
\end{figure}

A natural choice for non-locally encoding a qubit into the phase space of an oscillator is superpositions of macroscopically distinct coherent states, the so-called Schr\"{o}dinger cat states. Here we choose the states ${|\mathcal{C}_{\alpha}^{\pm}\rangle = (|+\alpha\rangle \pm |-\alpha\rangle)}/\sqrt{2}$ with average photon number $\bar{n}=|\alpha|^{2}$ and respectively even/odd photon number parity as the Z-basis of the encoded qubit (see Fig.~\ref{fig1}\bloch). The coherent states $\xpm$ are approximate X-basis vectors of the encoded qubit~\cite{SI}, and their distance in phase space ensures protection against any noise process that causes local displacements in phase space. Crucially, this leads to a suppression of phase flips that is exponential in the average photon number $\bar{n}$~\cite{Cochrane1999,Mirrahimi2014,SI}. In particular, photon loss, the usual noise process in an oscillator, cannot induce transitions between $\xpm$ because they are eigenstates of the annihilation operator $\hat{a}$.  This is however not the case for their superpositions, and a stochastic photon loss event corresponds to a bit-flip error on the encoded qubit: $\hat{a}\catpm=\alpha \catmp$. This flip also affects the parity-less states ${|\mathcal{C}_{\alpha}^{\mp i}\rangle = (|+\alpha\rangle \mp i |-\alpha\rangle) /\sqrt{2}}$ along the Y-axis of the encoded qubit. However, for a given single photon-loss rate $\kappa_a$, the bit-flip rate ${\approx2\bar{n}\kappa_a}$~\cite{Haroche2006,SI} increases only linearly with the photon number. A qubit with such ``biased noise'' is an important resource in fault-tolerant quantum computation~\cite{Aliferis2008, Tuckett2018}. Additional layers of error correction can then focus strongly on the remaining bit-flip error~\cite{Tuckett2018,Puri2019,Guillaud2019}. This significantly reduces the number of building blocks compared to conventional approaches that use qubits without such error protection. Another application of this type of protected qubit is to detect quantum errors in other encoded qubits without introducing additional errors, which is a crucial ingredient in fault-tolerant quantum computation~\cite{Puri2018a}.

In order for these applications to be practical, it is essential that operations can be performed on the protected qubit faster than the shortest decoherence time-scale, here the bit-flip time. Here we experimentally demonstrate such fast gate operations and quantum non-demolition single shot readout in a system that maintains phase-flip protection via the simultaneous stabilization of two opposite-phase coherent states.

Our approach is based on the application of a single-mode squeezing drive to a Kerr-nonlinear resonator~\cite{Puri2017}. In the frame rotating at the resonator frequency $\omega_{a}$, the system is described by the Hamiltonian

\begin{equation}
    \hat{H}\sub{cat}/\hbar =-K \hat{a}^{\dag 2}\hat{a}^{2}+\epsilon_{2}(\hat{a}^{\dag 2}+\hat{a}^{2}),
    \label{eq:H}
\end{equation}
where $K$ is the Kerr-nonlinearity and $\epsilon_{2}$ is the amplitude of the squeezing drive. Some intuition on this system can be gained from computing $\langle\hat{H}\sub{cat}\rangle$ as a function of classically treated phase-space coordinates. There are two stable extrema at $\pm\alpha=\pm \sqrt{\epsilon_2/K}$, as indicated by the markers in Fig.~\ref{fig1}\phasespace. They correspond to the lowest degenerate eigenstates of the quantum Hamiltonian~\cite{Puri2017,SI} and thus do not decay to vacuum. These eigenstates are separated from the rest of the spectrum by an energy gap ${E\sub{gap}/\hbar\approx 4 K\bar{n}}$~\cite{SI}, which sets the speed limit for operations and readout. The barrier between them prevents jumps along the X-axis of this ``Kerr-cat qubit''. However, if no squeezing drive is applied, $\hat{H}\sub{cat}$ reduces to the Hamiltonian of a simple anharmonic oscillator. This is similar to a superconducting transmon with anharmonicity $2K$, which is commonly used to encode a ``Fock qubit'' into the first two photon number states $\ket{0}$ and $\ket{1}$ (see Fig.~\ref{fig1}\blochfock). Its classical energy displays one single extremum in phase space harboring the quadrature expectation values of both states along the qubit X-axis without the protection of an energy barrier as shown in the right-hand panel of Fig.~\ref{fig1}\phasespace. By toggling the squeezing drive, a Kerr-nonlinear resonator can be tuned to implement either type of qubit.

\begin{figure*}
 \includegraphics[angle = 0, width = \figwidthWide]{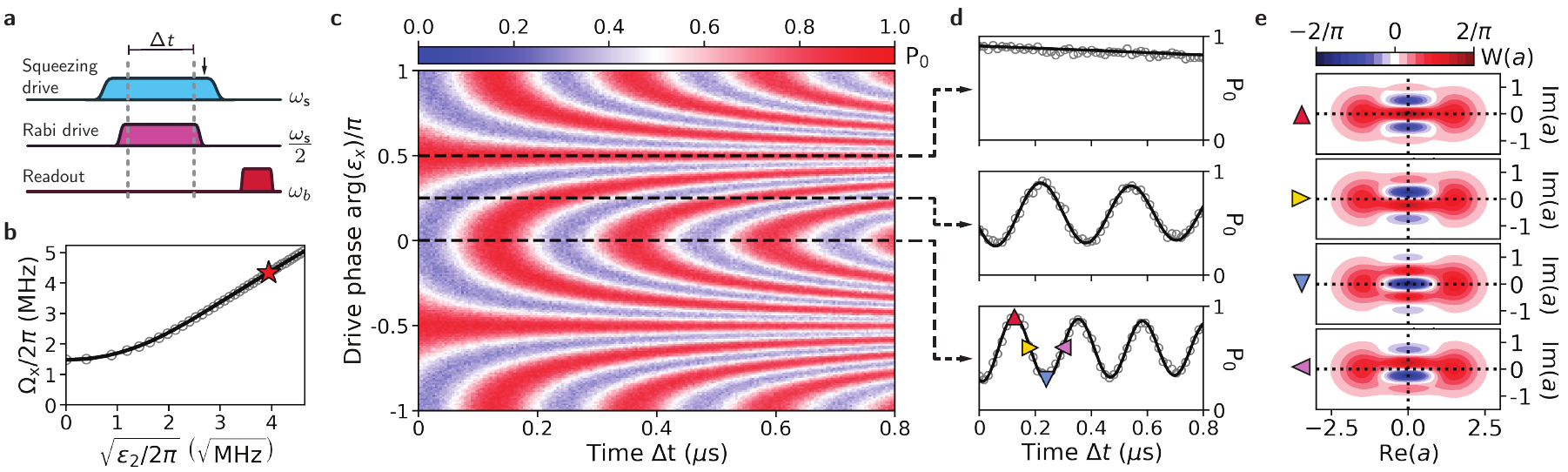}
 \caption{\label{fig2} \textbf{Rabi oscillations of the protected Kerr-cat-qubit. a}, Pulse sequence to perform the following functions: (i) initialize the Kerr-cat-qubit, (ii) drive Rabi oscillations for a varying time $\Delta t$, (iii) map onto the Fock qubit and perform dispersive readout. The frequencies of the single-mode squeezing drive, Rabi drive, and readout pulse are $\omega\sub{s}$, $\omega\sub{s}/2$, and $\omega_b$, respectively. A black arrow indicates the endpoint of the numerical simulations performed for \textbf{b},\textbf{d},\textbf{e}. \textbf{b}, Dependence of the Rabi frequency $\Omega_x$ on $\sqrt{\epsilon_2}$. Experimental data are open grey circles. The solid black line is a one-parameter fit to the data~\cite{SI} used to calibrate $\epsilon_2$. The red star indicates the condition $\epsilon_{2}/2\pi=\epstwofigtwo$ used for \textbf{c},\textbf{d},\textbf{e}. \textbf{c}, Dependence of the experimentally measured Rabi oscillations on time $\Delta t$ and on the phase of the Rabi drive. The color scale gives the ground state population of the Fock qubit ($P_0$) at the end of the sequence. Horizontal lines indicate the cuts shown in \textbf{d}. \textbf{d}, Cuts of \textbf{c} for three different Rabi-drive phases. Open grey circles are experimental data, black lines are simulation. Symbols in the bottom panel indicate the times for which the simulated oscillator state is shown in \textbf{e}. \textbf{e,} Simulated phase-space representation of the oscillator density matrix corresponding to the $\ket{+Z}$, $\ket{+Y}$, $\ket{-Z}$ and $\ket{-Y}$ states of the Kerr-cat qubit. The color scale gives the value of the Wigner function $W(a)$ as a function of the real and imaginary part of the phase-space coordinate $a$.}
\end{figure*}

Our experimental implementation consists of a superconducting nonlinear resonator placed inside a 3D-microwave cavity~\cite{SI} (see Fig.~\ref{fig1}\cavity). This is a setup commonly used in 3D-transmon qubits with a few key differences: Large capacitor pads help reduce the Kerr-nonlinearity of the resonator relative to a transmon and thus limit the drive strength $\epsilon_2$ necessary to reach appreciable coherent state amplitudes.
We orient the resonator such that its dipole moment is perpendicular to the electric field direction of the lowest-frequency cavity mode (${\omega_b/2\pi=\fres}$) to avoid strong hybridization despite the large pads, and reintroduce a small precisely-tuned coupling necessary for dispersive readout by slightly offsetting the pads (note $\delta$ in Fig.~\ref{fig1}\snailmon). Furthermore, instead of a single Josephson junction we employ a superconducting nonlinear asymmetric inductive element (SNAIL)~\cite{Frattini2017} (see Fig.~\ref{fig1}\snail). This makes the resonator flux-tunable and endows it with both third- and fourth-order nonlinearities. We use the former to generate single-mode squeezing by applying a coherent microwave drive ${\omega_s}$ at twice the resonator frequency, thus converting one drive photon into two resonator photons through three-wave mixing; the latter yields the required Kerr-nonlinearity~\cite{SI}. For the measurements presented in this work, we flux-tune our device to a frequency $\omega_a/2\pi=\fqubit$ and Kerr-nonlinearity ${K/2\pi=\kerr}$. At this frequency, the Fock qubit has an amplitude damping time $T_1=\bareTone$ and a transverse relaxation time $T_{2}=\bareTphi$.

The Fock qubit is employed for initialization and measurement of the Kerr-cat qubit during most experiments described in this work. This is possible, because the states $\ket{0},\catp$ ($\ket{1},\catm$) spanning the two Bloch spheres in Fig.~\ref{fig1}\bloch,\blochfock have the same even (odd) photon number parity, which is conserved by the system Hamiltonian in equation~(\ref{eq:H}). Consequently, for initialization, we prepare the Fock qubit in a given state and then map this state onto the Kerr-cat qubit by ramping on the squeezing drive with a hyperbolic-tangent profile over \tanhpump as sketched in Fig.~\ref{fig1}\phasespace. This time is chosen to be significantly longer than $1/2K$ to avoid leakage to higher excited states. After performing operations on the Kerr-cat qubit, we reverse the mapping and measure the state of the Fock qubit with dispersive readout. 

We now show that we indeed implement the Hamiltonian~(\ref{eq:H}), and thus initialize and stabilize a Kerr-cat qubit, by demonstrating unique features of Rabi-oscillations around the X-axis of its Bloch sphere. To this end, we apply an additional coherent drive $\epsilon_x \hat{a}^{\dag} +\epsilon_x^{*} \hat{a}$ at frequency $\omega_a=\omega_s/2$ to the system, as shown in Fig.~\ref{fig2}\pulseRabi. This lifts the degeneracy between the states $\xpm$ and therefore leads to oscillations with a Rabi frequency
\begin{equation}
    \Omega_x = \mathrm{Re}(4 \epsilon_x \alpha)
    \label{eq:rabi}
\end{equation}
between their superposition states along the purple circle shown in Fig.~\ref{fig1}\bloch. This picture is valid for large enough $\alpha$ and as long as the coherent states are not significantly displaced by the additional drive~\cite{SI}. Such a displacement is suppressed by the gap leading to the condition ${\epsilon_x \ll E\sub{gap}/\hbar}$~\cite{Puri2017,SI}. This sets the speed limit for Rabi oscillations on the Kerr-cat qubit significantly above the corresponding limit for the Fock qubit given by $2K$. Note that equation~(\ref{eq:rabi}) is different from the Rabi frequency of a Fock qubit in two ways: First, it depends on the amplitude of the squeezing drive through $\alpha \propto \sqrt{\epsilon_2}$. Second, it varies with the phase of the applied Rabi drive $\arg(\epsilon_x)$ (for simplicity we chose $\epsilon_2,\alpha\in \mathds{R}$).

We first focus on the effect of the squeezing drive on the Rabi frequency. We initialize in $\catp$ and apply a Rabi drive for a variable time $\Delta t$ with a constant amplitude and $\arg(\epsilon_x)=0$ as shown in Fig.~\ref{fig2}\pulseRabi. In addition, we vary $\epsilon_2$ and fit the Rabi frequencies of the oscillations in the measured Fock qubit $\ket{0}$-state population fraction at the end of the experiment (see Fig.~\ref{fig2}\Rabivsamp). For $\epsilon_2=0$ we are simply driving Rabi oscillations of the Fock qubit giving a direct calibration of $\epsilon_x / 2\pi =\epsx$~\cite{SI}. For large values of $\epsilon_2$ the Rabi frequency becomes a linear function of $\sqrt{\epsilon_2}$, confirming the theoretical prediction of equation~(\ref{eq:rabi}). We calibrate the abscissa of Fig.~\ref{fig2}\Rabivsamp by fitting the maximum Rabi frequency to a numerical simulation of the system dynamics, with the corresponding value of $\epsilon_2$ as the only free parameter~\cite{SI}. The result of this simulation extrapolated to all values of $\epsilon_2$ is shown as a solid black line.

We now turn to another unique feature of these Rabi oscillations by setting $\epsilon_2 / 2\pi =\epstwofigtwo$ and varying $\Delta t$ and $\arg(\epsilon_x)$. As expected, the measured oscillations shown in Fig.~\ref{fig2}\Rabivsphase are $\pi$-periodic in the phase of the Rabi drive. Three cuts through this data (indicated by dashed lines) are shown in Fig.~\ref{fig2}\cutsRabivsphase. The uppermost panel corresponds to a phase difference of $\pi/2$ between the coherent state amplitude and the Rabi drive, meaning that oscillations are suppressed. The middle and bottom panels at respective phase differences of $\pi/4$ and $0$ display increasing Rabi frequencies with the latter corresponding to the red star marker in Fig.~\ref{fig2}\Rabivsamp. We use the parameters obtained from Fig.~\ref{fig2}\Rabivsamp to perform a simulation of this experiment up to the point indicated by the black arrow in Fig.~\ref{fig2}\pulseRabi. The black line in the bottom panel of Fig.~\ref{fig2}\cutsRabivsphase is the result of this simulation, scaled to match the contrast of the data. The black lines in the two upper panels use the same scaling factor and are thus parameter-free predictions in good agreement with the measured data. Having benchmarked our simulation in this way, we use it to compute the full density matrix of the resonator at the time points indicated by markers in the figure. The corresponding Wigner function representations are shown in Fig.~\ref{fig2}\wigners. Apart from slight asymmetries due to the finite ramp time of the initial mapping pulse, they agree well with the expected $\ket{+Z}$, $\ket{+Y}$, $\ket{-Z}$, and $\ket{-Y}$ states (from top to bottom) of the Kerr-cat qubit.

\begin{figure}
 \includegraphics[angle = 0, width = \figwidth]{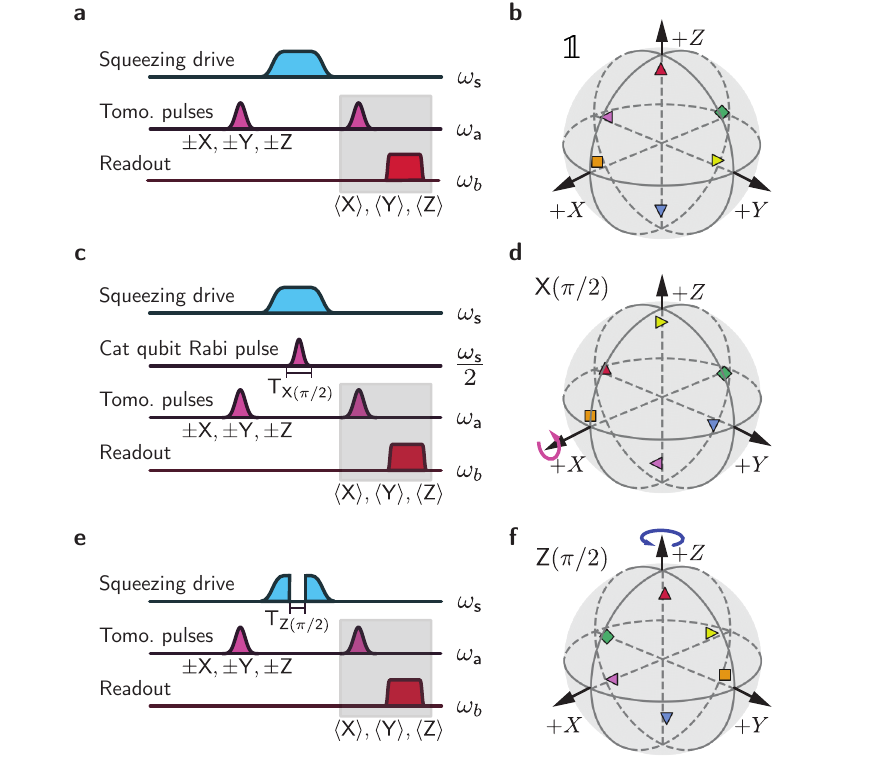}
 \caption{\label{fig3}  \textbf{Kerr-cat qubit gate process tomography. a}, \textbf{c}, \textbf{e}, Pulse sequences for process tomography of respectively: mapping between Fock qubit and Kerr-cat qubit (ideally $\mathds{1}$), mapping and $X(\pi/2)$ gate, mapping and $Z(\pi/2)$ gate. In each sequence, the Fock qubit is initialized on the $\pm X,\pm Y,\pm Z$ cardinal points of the Bloch sphere, the respective operation is performed, and the expectation values $\langle X \rangle, \langle Y \rangle, \langle Z \rangle$ are measured by a combination of Fock qubit pulses and dispersive readout (grey box). In \textbf{c}, $T_{X(\pi/2)}=\xgateT$ is the total duration of the Gaussian Rabi pulse applied to the Kerr-cat qubit. In \textbf{e}, $T_{Z(\pi/2)}=\zgateT$ is the duration for which the squeezing drive is switched off to perform the gate. \textbf{b}, \textbf{d}, \textbf{f} Process tomography results presented on Bloch sphere for respectively: mapping, $X(\pi/2)$ gate,  and $Z(\pi/2)$ gate. The expectation-value vector after the operation for an initialization on the ($+X,-X,+Y,-Y,+Z,-Z$) cardinal point is represented by an (orange square, green diamond, gold right-facing triangle, purple left-facing triangle, red upward-facing triangle, blue downward-facing triangle). The respective fidelities are $0.855\pm0.002$, $0.857\pm0.001$, and $0.811\pm0.001$.}
\end{figure}

Next, we characterize the mapping operation as well as a set of single-qubit gates on the Kerr-cat qubit by performing process tomography using the pulse sequences shown in Fig.~\ref{fig3}\pulseMap,\pulseX,\pulseZ. At the beginning of each sequence, the Kerr-cat qubit is initialized in one of the six states $\ket{\pm X},\ket{\pm Y},\ket{\pm Z}$. Each sequence ends with the reverse mapping and measurement of the resulting $\langle X \rangle$, $\langle Y \rangle$, and $\langle Z \rangle$ component of the Fock qubit state. In all subsequent experiments, the average photon number of the cat states is set to $\approx\nbarRest$ and frequency shifts induced by the squeezing drive are taken into account by setting $\omega\sub{s}/2=\tilde{\omega}_{a}$, where $\tilde{\omega}_{a}$ is the Stark-shifted resonator frequency~\cite{SI}.

The first operation we characterize in this way is the mapping between the Fock qubit and the Kerr-cat qubit. The corresponding pulse sequence is shown in Fig.~\ref{fig3}\pulseMap and the measured state-vectors are plotted on a Bloch sphere in Fig.~\ref{fig3}\blochMap. An estimate of the fidelity $\mathcal{F}\sub{map}\approx0.855\pm0.002$ ($\pm$ one standard deviation) is obtained by using the Pauli transfer matrix approach~\cite{SI, Chow2012a}. This number reflects the fidelities of the Fock qubit pulses as well as of the mapping itself.

Next, we turn to the pulse sequence shown in Fig.~\ref{fig3}\pulseX, which additionally performs an $X(\pi/2)$-gate on the Kerr-cat qubit. We apply a Gaussian pulse with a duration of $\xgateT$  and a maximum amplitude of $\epsilon\sub{x,\frac{\pi}{2}}=\epsxGate$. The process tomography data (see Fig.~\ref{fig3}\blochMap) displays the desired rotation around the X-axis with a fidelity of ${\mathcal{F}\sub{X(\pi/2)}\approx0.857\pm0.001}$. Comparing this value to $\mathcal{F}\sub{map}$ indicates that ${\mathcal{F}\sub{X(\pi/2)}}$ is mostly limited by the state initialization and mapping and not by the gate itself. From a complementary measurement~\cite{SI}, we estimate the infidelity due to over-rotation and decoherence during the gate operation to $\approx0.01$.

Since this operation is compatible with an arbitrary angle of rotation, only a $\pi/2$ rotation around the Z-axis is needed to reach any point on the Bloch sphere of the Kerr-cat qubit. Nominally, such a gate is incompatible with the stabilization as it could be used to go between the states $\xpm$. However, for $\epsilon_2=0$, the free evolution of the Kerr-Hamiltonian for a time $\pi/2K\approx\SI{37.3}{\nano \second}$ achieves the required operation~\cite{Yurke1988,Kirchmair2013,Puri2017}. We therefore implement a $Z(\pi/2)$ gate by abruptly setting the squeezing drive amplitude to zero for a duration $T_{Z(\pi/2)}=\zgateT$ and then switching it on again, as shown in Fig.~\ref{fig3}\pulseZ. The tomography data shown in Fig.~\ref{fig3}\blochZ gives a fidelity ${\mathcal{F}\sub{Z(\pi/2)}\approx0.811\pm0.001}$. We attribute the reduction of fidelity with respect to $\mathcal{F}\sub{map}$ to the difference between $\pi/2K$ and $T_{Z(\pi/2)}$ and to the finite rise time of the step function of $\approx\trise$, both of which are not limitations of our device, but of our experimental setup~\cite{SI}.

\begin{figure*}
 \includegraphics[angle = 0, width = \figwidthWide]{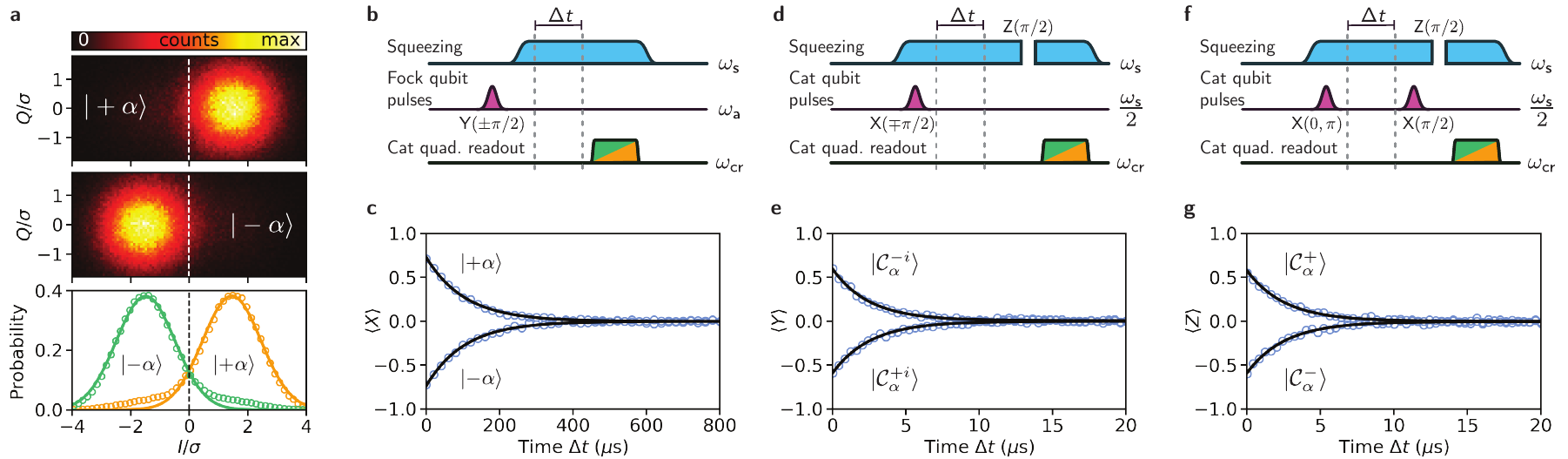}
\caption{\label{fig4} \textbf{Cat-quadrature readout and coherence times. a}, Top and middle panel: Histogram of $3\times10^{5}$ values of the demodulated and integrated cavity output field when performing cat-quadrature readout (see panel \textbf{b}) after preparation of either $\xp$ or $\xm$ as indicated. Bottom panel: Corresponding probability distribution along the $I$-quadrature. Open orange (green) circles show measured data for $\xp$ ($\xm$) and solid lines are Gaussian fits of width $\sigma$ used to scale the quadrature axes. Setting a threshold at $I/\sigma=0$ (dashed line) implements a direct single-shot readout of the Kerr-cat qubit along its $X$-axis with total fidelity $\mathcal{F} = \fid$. \textbf{b}, Cat-quadrature-readout pulse sequence for the measurements presented in \textbf{a} and \textbf{c}. After state initialization in $\xpm$, a pulse at frequency $\omega\sub{cr}=\omega_b-\omega\sub{s}/2$ is applied for a time $T\sub{cr}=\qsT$ converting the quadrature amplitude of nonlinear resonator hosting the Kerr-cat qubit to a drive on the readout cavity at $\omega_b$. The wait time $\Delta t$ is set to zero to obtain the results shown in \textbf{a} and varied to measure the coherence of the Kerr-cat-qubit $\langle X \rangle$-component shown in \textbf{c}. \textbf{c}, Kerr-cat qubit $\langle X \rangle$-component coherence. Open blue circles are data and solid black lines are single-exponential fits with decay times $\tau\sub{+X}={\catTphi\sdcoh}$ and $\tau\sub{-X}={\catTphim\sdcoh}$. \textbf{d}, Pulse sequence for coherence time measurements of the Kerr-cat qubit $\langle Y \rangle$-component shown in \textbf{e}. After initialization in $|\mathcal{C}_{\alpha}^{\mp i} \rangle$ (mapping $\ket{0}\rightarrow \catp$ and $X(\mp \pi/2)$ gate) and variable wait time $\Delta t$, a $Z(\pi/2)$ gate is performed followed by cat-quadrature readout. \textbf{e}, Kerr-cat qubit $\langle Y \rangle$-component coherence. Open blue circles are data and solid black lines are single-exponential fits with decay times ${\tau\sub{+Y}=\tauYp}$ and ${\tau\sub{-Y}=\tauYm}$. \textbf{f}, Pulse sequence for coherence time measurements of the Kerr-cat qubit $\langle Z \rangle$-component shown in \textbf{g}. After initialization in $\catpm$ (mapping $\ket{0}\rightarrow \catp$ and either $X(0)$ or $X(\pi)$ gate) and variable wait time $\Delta t$, a $X(\pi/2)$ gate and a $Z(\pi/2)$ gate are performed followed by cat-quadrature readout. \textbf{g}, Kerr-cat qubit $\langle Z \rangle$-component coherence. Open blue circles are data and solid black lines are single-exponential fits with decay times ${\tau\sub{+Z}=\tauZp}$ and ${\tau\sub{-Z}=\tauZm}$.}
\end{figure*}

So far, we have characterized the basic properties and gate operations of the Kerr-cat qubit by mapping back onto the Fock qubit and using the well-understood dispersive readout method. This readout, however, destroys the state of the Kerr-cat qubit. We now demonstrate an entirely new way to perform a quantum non-demolition measurement on the X-component of the stabilized Kerr-cat qubit, which we call the ``cat-quadrature readout''. To that end, we apply an additional drive at frequency $\omega\sub{cr}=\omega_b-\omega_s/2$ which, through the three-wave mixing capability of our system, generates a frequency-converting interaction between the nonlinear resonator and the readout cavity. In the frame rotating at both $\omega_s/2=\tilde{\omega}_a$ and $\omega_b$, this adds the following term to equation~(\ref{eq:H}):
\begin{equation}
    H\sub{cr}/\hbar = ig\sub{cr}(\hat{a} \hat{b}^{\dag}- \hat{a}^{\dag} \hat{b})
    \label{eq:qr}
\end{equation}
Here, $\hat{b}$ is the annihilation operator of the cavity field and $g\sub{cr}/2 \pi=\g$ is the independently measured coupling strength~\cite{SI}. For a quadrature expectation value ${\langle \hat{a} + \hat{a}^{\dag} \rangle /2 = \pm \alpha}$ in the nonlinear resonator, this causes an effective coherent drive on the cavity, populating it with a coherent state $|\pm \beta\rangle$~\cite{SI}, and projects the Kerr-cat qubit onto the corresponding state $\xpm$ along its X-axis. We gain information about the result of this projection by performing a heterodyne measurement on the outgoing cavity field (see Fig.~\ref{fig4}\qswitch).

We characterize the fidelity of this readout by first initializing the Kerr-cat qubit along its X-axis and then applying a cat-quadrature readout pulse for a time $T\sub{cr}=\qsT$ as shown in Fig.~\ref{fig4}\pulseQswitch.
Two histograms of $3\times 10^{5}$ measurements of the demodulated and integrated outgoing cavity field~\cite{SI} are shown in the top and bottom panel of Fig.~\ref{fig4}\qswitch for initialization in respectively $\xp$ or $\xm$. The probability distributions along the I-quadrature are shown as open orange and green circles in the bottom panel of Fig.~\ref{fig4}\qswitch while solid lines correspond to Gaussian fits of width $\sigma$, used to scale the axes. The separation of the two histograms is large enough to implement a single-shot readout by setting a threshold at $I/\sigma=0$. We define the total readout fidelity as $\mathcal{F}=1-p(-\alpha|+\alpha)-p(+\alpha|-\alpha)=\fid$, where $p(-\alpha|+\alpha)=\probmp$ ($p(+\alpha|-\alpha)=\probpm$) is the probability of measuring the qubit in $\xm$ ($\xp$) after initialization in $\xp$ ($\xm$). This value is a lower bound including errors in state preparation caused by the thermal population of the Fock qubit $\ket{1}$-state (contributing an infidelity of $\approx2\times\SI{4}{\percent}=\SI{8}{\percent}$) and imperfections during the initial Fock qubit pulse and mapping. Finally, we characterize the quantum-non-demolition aspect of this readout by evaluating $\mathcal{Q} = (p(+\alpha | +\alpha) + p(-\alpha | -\alpha))/2= \qndness$, where $p(i|i)$ is the probability of obtaining the measurement outcome $i$ in two successive measurements~\cite{SI}.

We now use this cat-quadrature readout to investigate the coherence times of the Kerr-cat qubit. We apply the same pulse sequence as described above but insert a variable wait time $\Delta t$ before the readout (see Fig.~\ref{fig4}\pulseQswitch) to measure the phase-flip time. The decay of the measured $\langle X \rangle$-component for either initial state is shown in Fig.~\ref{fig4}\tauX. We fit the data to a single-exponential decay with characteristic times $\tau\sub{+X}={\catTphi\sdcoh}$ and $\tau\sub{-X}={\catTphim\sdcoh}$. We additionally confirm this result through the complementary measurement which maps back onto the Fock qubit and performs dispersive readout (see supplementary material~\cite{SI}).

Similarly, the coherence times of both the $\langle Y \rangle$ and $\langle Z \rangle$ components are measured using the cat-quadrature readout, but employing only operations on the Kerr-cat qubit after an initial mapping operation from the $\ket{0}$-state of the Fock qubit to the cat state $\catp$ (see Fig.~\ref{fig4}\pulseTauY,\pulseTauZ). The resulting decay curves for both components are displayed in Fig.~\ref{fig4}\tauY,\tauZ. Single-exponential fits of the data yield the decay times ${\tau\sub{+Y}=\tauYp}$, ${\tau\sub{-Y}=\tauYm}$, ${\tau\sub{+Z}=\tauZp}$ and ${\tau\sub{-Z}=\tauZm}$. These values are slightly smaller than the predicted bit-flip time due to photon loss $\tau\sub{loss}=T\sub{1}/2\bar{n}=\SI{2.98}{\micro\second}$. We expect that an additional contribution from photon gain processes plays a role in this reduction~\cite{SI}.

Our results demonstrate a thirty-fold increase in the phase-flip time of the protected Kerr-cat qubit with respect to the Fock qubit. Crucially, we also show that we can perform a full set of single-qubit gates on the protected qubit on timescales that are significantly shorter than its bit-flip time. Although the measured gate fidelities are limited by state preparation and measurement errors, an upper bound of the error rate due to decoherence during the gate operations is given by: ${T\sub{X(\pi/2)}/\tau\sub{+Y}\approx T\sub{Z(\pi/2)}/\tau\sub{+Y}<2\times10^{-2}}$. The combination of error protection, fast gates, and single-shot readout opens the door to using stabilized Schr\"{o}dinger cat states as physical qubits in a future quantum computer. The simplicity of our implementation provides a straightforward path to coupling several Kerr-cat qubits, even within the same cavity, and demonstrating operations between them. In particular, our qubit permits a noise-bias-preserving controlled-NOT gate~\cite{Puri2019} which would be impossible with ordinary qubits~\cite{Guillaud2019}. This will require further improvements in device performance as such operations are limited by the bit-flip time, which in the current system is limited by losses due to the copper cavity. Implementing a magnetic flux bias in a superconducting enclosure~\cite{Gargiulo2018} will help to relax this constraint and additionally increase the achievable gate and readout fidelities. Kerr-cat qubits could then also be applied as auxiliary systems for fault-tolerant error detection on other logical qubits~\cite{Puri2018a}.

Another effect requiring further investigation is the limit on the phase-flip time. Measurements performed at other flux points with different strengths of the third- and fourth-order nonlinearities indicate that coherence decreases when stronger drives have to be applied to the system to reach similar photon numbers. Similarly, an increase in photon number beyond the operating point chosen in this work leads to a decrease in coherence. This decrease is probably related to heating effects associated with the strong driving of Josephson-junction devices~\cite{Sank2016a,Lescanne2019b} causing leakage to higher excited states outside of the Kerr-cat-qubit encoding. It is likely that such leakage can be counter-acted through controlled two-photon dissipation back towards the states of the Kerr-cat qubit~\cite{Puri2018a}. Such a dissipation-based approach to stabilization can be effective in achieving strongly biased noise~\cite{Lescanne2019}, but quantum operations that are much faster than all coherence timescales of the encoded qubit remain difficult to achieve~\cite{Touzard2018a}. The optimal solution should be to rely on a combination of two-photon dissipation and Kerr for phase-flip suppression and on the Kerr-effect for high gate speeds.

In addition to its applications in fault-tolerant quantum computation~\cite{Puri2017,Puri2018a,Puri2019,Goto2016a}, our system extends the understanding of bistability in parametrically driven Kerr-nonlinear oscillators from the classical regime where losses dominate over the nonlinearity~\cite{Dykman1998a,Wustmann2013b,Siddiqi2005,Wilson2010} to the inverse case where quantum states can be generated~\cite{Yurke1988,Kirchmair2013,Zhang2017,Wang2019a}. The present work demonstrates for the first time long-lived quantum superpositions of macroscopically distinct classical states due to bistability in a nonlinear oscillator. Such states could shed light on the quantum-classical transition~\cite{Zurek2003} and can be useful in weak force measurements~\cite{Munro2002a}.  Networks of coupled bistable oscillators can be mapped onto Ising spins and used to investigate non-equilibrium quantum phase transitions~\cite{Dykman2018} or to solve combinatorial optimization problems~\cite{Puri2017e,Goto2016b,Marandi2014}. These examples suggest that our Kerr-cat system is likely to be applied both to quantum computation and for the investigation of fundamental quantum effects.

\bibliographystyle{apsrev}
\bibliography{bibliography_updated_arxiv_final_version}

\begin{thebibliography}{50}
\expandafter\ifx\csname natexlab\endcsname\relax\def\natexlab#1{#1}\fi
\expandafter\ifx\csname bibnamefont\endcsname\relax
  \def\bibnamefont#1{#1}\fi
\expandafter\ifx\csname bibfnamefont\endcsname\relax
  \def\bibfnamefont#1{#1}\fi
\expandafter\ifx\csname citenamefont\endcsname\relax
  \def\citenamefont#1{#1}\fi
\expandafter\ifx\csname url\endcsname\relax
  \def\url#1{\texttt{#1}}\fi
\expandafter\ifx\csname urlprefix\endcsname\relax\def\urlprefix{URL }\fi
\providecommand{\bibinfo}[2]{#2}
\providecommand{\eprint}[2][]{\url{#2}}

\bibitem[{\citenamefont{Cochrane et~al.}(1999)\citenamefont{Cochrane, Milburn,
  and Munro}}]{Cochrane1999}
\bibinfo{author}{\bibfnamefont{P.~T.} \bibnamefont{Cochrane}},
  \bibinfo{author}{\bibfnamefont{G.~J.} \bibnamefont{Milburn}},
  \bibnamefont{and} \bibinfo{author}{\bibfnamefont{W.~J.} \bibnamefont{Munro}},
  \bibinfo{journal}{Phys. Rev. A} \textbf{\bibinfo{volume}{59}},
  \bibinfo{pages}{2631} (\bibinfo{year}{1999}).

\bibitem[{\citenamefont{Mirrahimi et~al.}(2014)\citenamefont{Mirrahimi,
  Leghtas, Albert, Touzard, Schoelkopf, Jiang, and Devoret}}]{Mirrahimi2014}
\bibinfo{author}{\bibfnamefont{M.}~\bibnamefont{Mirrahimi}},
  \bibinfo{author}{\bibfnamefont{Z.}~\bibnamefont{Leghtas}},
  \bibinfo{author}{\bibfnamefont{V.~V.} \bibnamefont{Albert}},
  \bibinfo{author}{\bibfnamefont{S.}~\bibnamefont{Touzard}},
  \bibinfo{author}{\bibfnamefont{R.~J.} \bibnamefont{Schoelkopf}},
  \bibinfo{author}{\bibfnamefont{L.}~\bibnamefont{Jiang}}, \bibnamefont{and}
  \bibinfo{author}{\bibfnamefont{M.~H.} \bibnamefont{Devoret}},
  \bibinfo{journal}{New J. Phys.} \textbf{\bibinfo{volume}{16}},
  \bibinfo{pages}{045014} (\bibinfo{year}{2014}).

\bibitem[{\citenamefont{Milburn and Holmes}(1991)}]{Milburn1991}
\bibinfo{author}{\bibfnamefont{G.~J.} \bibnamefont{Milburn}} \bibnamefont{and}
  \bibinfo{author}{\bibfnamefont{C.~A.} \bibnamefont{Holmes}},
  \bibinfo{journal}{Phys. Rev. A} \textbf{\bibinfo{volume}{44}},
  \bibinfo{pages}{4704} (\bibinfo{year}{1991}).

\bibitem[{\citenamefont{Puri et~al.}(2017{\natexlab{a}})\citenamefont{Puri,
  Boutin, and Blais}}]{Puri2017}
\bibinfo{author}{\bibfnamefont{S.}~\bibnamefont{Puri}},
  \bibinfo{author}{\bibfnamefont{S.}~\bibnamefont{Boutin}}, \bibnamefont{and}
  \bibinfo{author}{\bibfnamefont{A.}~\bibnamefont{Blais}},
  \bibinfo{journal}{Npj Quantum Inf.} \textbf{\bibinfo{volume}{3}},
  \bibinfo{pages}{18} (\bibinfo{year}{2017}{\natexlab{a}}).

\bibitem[{\citenamefont{Puri et~al.}(2019{\natexlab{a}})\citenamefont{Puri,
  Grimm, Campagne-Ibarcq, Eickbusch, Noh, Roberts, Jiang, Mirrahimi, Devoret,
  and Girvin}}]{Puri2018a}
\bibinfo{author}{\bibfnamefont{S.}~\bibnamefont{Puri}},
  \bibinfo{author}{\bibfnamefont{A.}~\bibnamefont{Grimm}},
  \bibinfo{author}{\bibfnamefont{P.}~\bibnamefont{Campagne-Ibarcq}},
  \bibinfo{author}{\bibfnamefont{A.}~\bibnamefont{Eickbusch}},
  \bibinfo{author}{\bibfnamefont{K.}~\bibnamefont{Noh}},
  \bibinfo{author}{\bibfnamefont{G.}~\bibnamefont{Roberts}},
  \bibinfo{author}{\bibfnamefont{L.}~\bibnamefont{Jiang}},
  \bibinfo{author}{\bibfnamefont{M.}~\bibnamefont{Mirrahimi}},
  \bibinfo{author}{\bibfnamefont{M.~H.} \bibnamefont{Devoret}},
  \bibnamefont{and} \bibinfo{author}{\bibfnamefont{S.~M.}
  \bibnamefont{Girvin}}, \bibinfo{journal}{Phys. Rev. X}
  \textbf{\bibinfo{volume}{9}}, \bibinfo{pages}{041009}
  (\bibinfo{year}{2019}{\natexlab{a}}).

\bibitem[{\citenamefont{Puri et~al.}(2019{\natexlab{b}})\citenamefont{Puri,
  St-Jean, Gross, Grimm, Frattini, Iyer, Krishna, Touzard, Jiang, Blais
  et~al.}}]{Puri2019}
\bibinfo{author}{\bibfnamefont{S.}~\bibnamefont{Puri}},
  \bibinfo{author}{\bibfnamefont{L.}~\bibnamefont{St-Jean}},
  \bibinfo{author}{\bibfnamefont{J.~A.} \bibnamefont{Gross}},
  \bibinfo{author}{\bibfnamefont{A.}~\bibnamefont{Grimm}},
  \bibinfo{author}{\bibfnamefont{N.~E.} \bibnamefont{Frattini}},
  \bibinfo{author}{\bibfnamefont{P.~S.} \bibnamefont{Iyer}},
  \bibinfo{author}{\bibfnamefont{A.}~\bibnamefont{Krishna}},
  \bibinfo{author}{\bibfnamefont{S.}~\bibnamefont{Touzard}},
  \bibinfo{author}{\bibfnamefont{L.}~\bibnamefont{Jiang}},
  \bibinfo{author}{\bibfnamefont{A.}~\bibnamefont{Blais}},
  \bibnamefont{et~al.}, \bibinfo{journal}{ArXiv190500450}
  (\bibinfo{year}{2019}{\natexlab{b}}).

\bibitem[{\citenamefont{Guillaud and Mirrahimi}(2019)}]{Guillaud2019}
\bibinfo{author}{\bibfnamefont{J.}~\bibnamefont{Guillaud}} \bibnamefont{and}
  \bibinfo{author}{\bibfnamefont{M.}~\bibnamefont{Mirrahimi}},
  \bibinfo{journal}{Phys. Rev. X} \textbf{\bibinfo{volume}{9}},
  \bibinfo{pages}{041053} (\bibinfo{year}{2019}).

\bibitem[{\citenamefont{Goto}(2016{\natexlab{a}})}]{Goto2016b}
\bibinfo{author}{\bibfnamefont{H.}~\bibnamefont{Goto}}, \bibinfo{journal}{Sci.
  Rep.} \textbf{\bibinfo{volume}{6}}, \bibinfo{pages}{21686}
  (\bibinfo{year}{2016}{\natexlab{a}}).

\bibitem[{\citenamefont{Shor}(1995)}]{Shor1995}
\bibinfo{author}{\bibfnamefont{P.~W.} \bibnamefont{Shor}},
  \bibinfo{journal}{Phys. Rev. A} \textbf{\bibinfo{volume}{52}},
  \bibinfo{pages}{R2493} (\bibinfo{year}{1995}).

\bibitem[{\citenamefont{Kitaev}(2003)}]{Kitaev2003}
\bibinfo{author}{\bibfnamefont{A.~Y.} \bibnamefont{Kitaev}},
  \bibinfo{journal}{Annals of Physics} \textbf{\bibinfo{volume}{303}},
  \bibinfo{pages}{2} (\bibinfo{year}{2003}).

\bibitem[{\citenamefont{Oreg et~al.}(2010)\citenamefont{Oreg, Refael, and von
  Oppen}}]{Oreg2010}
\bibinfo{author}{\bibfnamefont{Y.}~\bibnamefont{Oreg}},
  \bibinfo{author}{\bibfnamefont{G.}~\bibnamefont{Refael}}, \bibnamefont{and}
  \bibinfo{author}{\bibfnamefont{F.}~\bibnamefont{von Oppen}},
  \bibinfo{journal}{Phys. Rev. Lett.} \textbf{\bibinfo{volume}{105}},
  \bibinfo{pages}{177002} (\bibinfo{year}{2010}).

\bibitem[{\citenamefont{Lutchyn et~al.}(2010)\citenamefont{Lutchyn, Sau, and
  Das~Sarma}}]{Lutchyn2010}
\bibinfo{author}{\bibfnamefont{R.~M.} \bibnamefont{Lutchyn}},
  \bibinfo{author}{\bibfnamefont{J.~D.} \bibnamefont{Sau}}, \bibnamefont{and}
  \bibinfo{author}{\bibfnamefont{S.}~\bibnamefont{Das~Sarma}},
  \bibinfo{journal}{Phys. Rev. Lett.} \textbf{\bibinfo{volume}{105}},
  \bibinfo{pages}{077001} (\bibinfo{year}{2010}).

\bibitem[{\citenamefont{Fowler et~al.}(2012)\citenamefont{Fowler, Mariantoni,
  Martinis, and Cleland}}]{Fowler2012a}
\bibinfo{author}{\bibfnamefont{A.~G.} \bibnamefont{Fowler}},
  \bibinfo{author}{\bibfnamefont{M.}~\bibnamefont{Mariantoni}},
  \bibinfo{author}{\bibfnamefont{J.~M.} \bibnamefont{Martinis}},
  \bibnamefont{and} \bibinfo{author}{\bibfnamefont{A.~N.}
  \bibnamefont{Cleland}}, \bibinfo{journal}{Phys. Rev. A}
  \textbf{\bibinfo{volume}{86}}, \bibinfo{pages}{032324}
  (\bibinfo{year}{2012}).

\bibitem[{\citenamefont{Gottesman et~al.}(2001)\citenamefont{Gottesman, Kitaev,
  and Preskill}}]{Gottesman2001a}
\bibinfo{author}{\bibfnamefont{D.}~\bibnamefont{Gottesman}},
  \bibinfo{author}{\bibfnamefont{A.}~\bibnamefont{Kitaev}}, \bibnamefont{and}
  \bibinfo{author}{\bibfnamefont{J.}~\bibnamefont{Preskill}},
  \bibinfo{journal}{Phys. Rev. A} \textbf{\bibinfo{volume}{64}},
  \bibinfo{pages}{012310} (\bibinfo{year}{2001}).

\bibitem[{\citenamefont{Vuillot et~al.}(2019)\citenamefont{Vuillot, Asasi,
  Wang, Pryadko, and Terhal}}]{Vuillot2019}
\bibinfo{author}{\bibfnamefont{C.}~\bibnamefont{Vuillot}},
  \bibinfo{author}{\bibfnamefont{H.}~\bibnamefont{Asasi}},
  \bibinfo{author}{\bibfnamefont{Y.}~\bibnamefont{Wang}},
  \bibinfo{author}{\bibfnamefont{L.~P.} \bibnamefont{Pryadko}},
  \bibnamefont{and} \bibinfo{author}{\bibfnamefont{B.~M.}
  \bibnamefont{Terhal}}, \bibinfo{journal}{Phys. Rev. A}
  \textbf{\bibinfo{volume}{99}}, \bibinfo{pages}{032344}
  (\bibinfo{year}{2019}).

\bibitem[{SI()}]{SI}
\bibinfo{note}{See Supplementary Material}.

\bibitem[{\citenamefont{Haroche and Raimond}(2006)}]{Haroche2006}
\bibinfo{author}{\bibfnamefont{S.}~\bibnamefont{Haroche}} \bibnamefont{and}
  \bibinfo{author}{\bibfnamefont{J.-M.} \bibnamefont{Raimond}},
  \emph{\bibinfo{title}{Exploring the Quantum: Atoms, Cavities and Photons}}
  (\bibinfo{publisher}{{Oxford University Press}}, \bibinfo{year}{2006}).

\bibitem[{\citenamefont{Aliferis and Preskill}(2008)}]{Aliferis2008}
\bibinfo{author}{\bibfnamefont{P.}~\bibnamefont{Aliferis}} \bibnamefont{and}
  \bibinfo{author}{\bibfnamefont{J.}~\bibnamefont{Preskill}},
  \bibinfo{journal}{Phys. Rev. A} \textbf{\bibinfo{volume}{78}},
  \bibinfo{pages}{052331} (\bibinfo{year}{2008}).

\bibitem[{\citenamefont{Tuckett et~al.}(2018)\citenamefont{Tuckett, Bartlett,
  and Flammia}}]{Tuckett2018}
\bibinfo{author}{\bibfnamefont{D.~K.} \bibnamefont{Tuckett}},
  \bibinfo{author}{\bibfnamefont{S.~D.} \bibnamefont{Bartlett}},
  \bibnamefont{and} \bibinfo{author}{\bibfnamefont{S.~T.}
  \bibnamefont{Flammia}}, \bibinfo{journal}{Phys. Rev. Lett.}
  \textbf{\bibinfo{volume}{120}}, \bibinfo{pages}{050505}
  (\bibinfo{year}{2018}).

\bibitem[{\citenamefont{Frattini et~al.}(2017)\citenamefont{Frattini, Vool,
  Shankar, Narla, Sliwa, and Devoret}}]{Frattini2017}
\bibinfo{author}{\bibfnamefont{N.~E.} \bibnamefont{Frattini}},
  \bibinfo{author}{\bibfnamefont{U.}~\bibnamefont{Vool}},
  \bibinfo{author}{\bibfnamefont{S.}~\bibnamefont{Shankar}},
  \bibinfo{author}{\bibfnamefont{A.}~\bibnamefont{Narla}},
  \bibinfo{author}{\bibfnamefont{K.~M.} \bibnamefont{Sliwa}}, \bibnamefont{and}
  \bibinfo{author}{\bibfnamefont{M.~H.} \bibnamefont{Devoret}},
  \bibinfo{journal}{Appl. Phys. Lett.} \textbf{\bibinfo{volume}{110}},
  \bibinfo{pages}{222603} (\bibinfo{year}{2017}).

\bibitem[{\citenamefont{Chow et~al.}(2012)\citenamefont{Chow, Gambetta,
  C\'{o}rcoles, Merkel, Smolin, Rigetti, Poletto, Keefe, Rothwell, Rozen
  et~al.}}]{Chow2012a}
\bibinfo{author}{\bibfnamefont{J.~M.} \bibnamefont{Chow}},
  \bibinfo{author}{\bibfnamefont{J.~M.} \bibnamefont{Gambetta}},
  \bibinfo{author}{\bibfnamefont{A.~D.} \bibnamefont{C\'{o}rcoles}},
  \bibinfo{author}{\bibfnamefont{S.~T.} \bibnamefont{Merkel}},
  \bibinfo{author}{\bibfnamefont{J.~A.} \bibnamefont{Smolin}},
  \bibinfo{author}{\bibfnamefont{C.}~\bibnamefont{Rigetti}},
  \bibinfo{author}{\bibfnamefont{S.}~\bibnamefont{Poletto}},
  \bibinfo{author}{\bibfnamefont{G.~A.} \bibnamefont{Keefe}},
  \bibinfo{author}{\bibfnamefont{M.~B.} \bibnamefont{Rothwell}},
  \bibinfo{author}{\bibfnamefont{J.~R.} \bibnamefont{Rozen}},
  \bibnamefont{et~al.}, \bibinfo{journal}{Phys. Rev. Lett.}
  \textbf{\bibinfo{volume}{109}}, \bibinfo{pages}{060501}
  (\bibinfo{year}{2012}).

\bibitem[{\citenamefont{Yurke and Stoler}(1988)}]{Yurke1988}
\bibinfo{author}{\bibfnamefont{B.}~\bibnamefont{Yurke}} \bibnamefont{and}
  \bibinfo{author}{\bibfnamefont{D.}~\bibnamefont{Stoler}},
  \bibinfo{journal}{Physica B+C} \textbf{\bibinfo{volume}{151}},
  \bibinfo{pages}{298} (\bibinfo{year}{1988}).

\bibitem[{\citenamefont{Kirchmair et~al.}(2013)\citenamefont{Kirchmair,
  Vlastakis, Leghtas, Nigg, Paik, Ginossar, Mirrahimi, Frunzio, Girvin, and
  Schoelkopf}}]{Kirchmair2013}
\bibinfo{author}{\bibfnamefont{G.}~\bibnamefont{Kirchmair}},
  \bibinfo{author}{\bibfnamefont{B.}~\bibnamefont{Vlastakis}},
  \bibinfo{author}{\bibfnamefont{Z.}~\bibnamefont{Leghtas}},
  \bibinfo{author}{\bibfnamefont{S.~E.} \bibnamefont{Nigg}},
  \bibinfo{author}{\bibfnamefont{H.}~\bibnamefont{Paik}},
  \bibinfo{author}{\bibfnamefont{E.}~\bibnamefont{Ginossar}},
  \bibinfo{author}{\bibfnamefont{M.}~\bibnamefont{Mirrahimi}},
  \bibinfo{author}{\bibfnamefont{L.}~\bibnamefont{Frunzio}},
  \bibinfo{author}{\bibfnamefont{S.~M.} \bibnamefont{Girvin}},
  \bibnamefont{and} \bibinfo{author}{\bibfnamefont{R.~J.}
  \bibnamefont{Schoelkopf}}, \bibinfo{journal}{Nature}
  \textbf{\bibinfo{volume}{495}}, \bibinfo{pages}{205} (\bibinfo{year}{2013}).

\bibitem[{\citenamefont{Gargiulo et~al.}(2018)\citenamefont{Gargiulo, Oleschko,
  Prat-Camps, Zanner, and Kirchmair}}]{Gargiulo2018}
\bibinfo{author}{\bibfnamefont{O.}~\bibnamefont{Gargiulo}},
  \bibinfo{author}{\bibfnamefont{S.}~\bibnamefont{Oleschko}},
  \bibinfo{author}{\bibfnamefont{J.}~\bibnamefont{Prat-Camps}},
  \bibinfo{author}{\bibfnamefont{M.}~\bibnamefont{Zanner}}, \bibnamefont{and}
  \bibinfo{author}{\bibfnamefont{G.}~\bibnamefont{Kirchmair}},
  \bibinfo{journal}{ArXiv181110875}  (\bibinfo{year}{2018}).

\bibitem[{\citenamefont{Sank et~al.}(2016)\citenamefont{Sank, Chen, Khezri,
  Kelly, Barends, Campbell, Chen, Chiaro, Dunsworth, Fowler
  et~al.}}]{Sank2016a}
\bibinfo{author}{\bibfnamefont{D.}~\bibnamefont{Sank}},
  \bibinfo{author}{\bibfnamefont{Z.}~\bibnamefont{Chen}},
  \bibinfo{author}{\bibfnamefont{M.}~\bibnamefont{Khezri}},
  \bibinfo{author}{\bibfnamefont{J.}~\bibnamefont{Kelly}},
  \bibinfo{author}{\bibfnamefont{R.}~\bibnamefont{Barends}},
  \bibinfo{author}{\bibfnamefont{B.}~\bibnamefont{Campbell}},
  \bibinfo{author}{\bibfnamefont{Y.}~\bibnamefont{Chen}},
  \bibinfo{author}{\bibfnamefont{B.}~\bibnamefont{Chiaro}},
  \bibinfo{author}{\bibfnamefont{A.}~\bibnamefont{Dunsworth}},
  \bibinfo{author}{\bibfnamefont{A.}~\bibnamefont{Fowler}},
  \bibnamefont{et~al.}, \bibinfo{journal}{Phys. Rev. Lett.}
  \textbf{\bibinfo{volume}{117}}, \bibinfo{pages}{190503}
  (\bibinfo{year}{2016}).

\bibitem[{\citenamefont{Lescanne et~al.}(2019)\citenamefont{Lescanne, Verney,
  Ficheux, Devoret, Huard, Mirrahimi, and Leghtas}}]{Lescanne2019b}
\bibinfo{author}{\bibfnamefont{R.}~\bibnamefont{Lescanne}},
  \bibinfo{author}{\bibfnamefont{L.}~\bibnamefont{Verney}},
  \bibinfo{author}{\bibfnamefont{Q.}~\bibnamefont{Ficheux}},
  \bibinfo{author}{\bibfnamefont{M.~H.} \bibnamefont{Devoret}},
  \bibinfo{author}{\bibfnamefont{B.}~\bibnamefont{Huard}},
  \bibinfo{author}{\bibfnamefont{M.}~\bibnamefont{Mirrahimi}},
  \bibnamefont{and} \bibinfo{author}{\bibfnamefont{Z.}~\bibnamefont{Leghtas}},
  \bibinfo{journal}{Phys. Rev. Applied} \textbf{\bibinfo{volume}{11}},
  \bibinfo{pages}{014030} (\bibinfo{year}{2019}).

\bibitem[{\citenamefont{Lescanne et~al.}(2020)\citenamefont{Lescanne, Villiers,
  Peronnin, Sarlette, Delbecq, Huard, Kontos, Mirrahimi, and
  Leghtas}}]{Lescanne2019}
\bibinfo{author}{\bibfnamefont{R.}~\bibnamefont{Lescanne}},
  \bibinfo{author}{\bibfnamefont{M.}~\bibnamefont{Villiers}},
  \bibinfo{author}{\bibfnamefont{T.}~\bibnamefont{Peronnin}},
  \bibinfo{author}{\bibfnamefont{A.}~\bibnamefont{Sarlette}},
  \bibinfo{author}{\bibfnamefont{M.}~\bibnamefont{Delbecq}},
  \bibinfo{author}{\bibfnamefont{B.}~\bibnamefont{Huard}},
  \bibinfo{author}{\bibfnamefont{T.}~\bibnamefont{Kontos}},
  \bibinfo{author}{\bibfnamefont{M.}~\bibnamefont{Mirrahimi}},
  \bibnamefont{and} \bibinfo{author}{\bibfnamefont{Z.}~\bibnamefont{Leghtas}},
  \bibinfo{journal}{Nature Physics} \textbf{\bibinfo{volume}{16}},
  \bibinfo{pages}{509–513} (\bibinfo{year}{2020}).

\bibitem[{\citenamefont{Touzard et~al.}(2018)\citenamefont{Touzard, Grimm,
  Leghtas, Mundhada, Reinhold, Axline, Reagor, Chou, Blumoff, Sliwa
  et~al.}}]{Touzard2018a}
\bibinfo{author}{\bibfnamefont{S.}~\bibnamefont{Touzard}},
  \bibinfo{author}{\bibfnamefont{A.}~\bibnamefont{Grimm}},
  \bibinfo{author}{\bibfnamefont{Z.}~\bibnamefont{Leghtas}},
  \bibinfo{author}{\bibfnamefont{S.}~\bibnamefont{Mundhada}},
  \bibinfo{author}{\bibfnamefont{P.}~\bibnamefont{Reinhold}},
  \bibinfo{author}{\bibfnamefont{C.}~\bibnamefont{Axline}},
  \bibinfo{author}{\bibfnamefont{M.}~\bibnamefont{Reagor}},
  \bibinfo{author}{\bibfnamefont{K.}~\bibnamefont{Chou}},
  \bibinfo{author}{\bibfnamefont{J.}~\bibnamefont{Blumoff}},
  \bibinfo{author}{\bibfnamefont{K.}~\bibnamefont{Sliwa}},
  \bibnamefont{et~al.}, \bibinfo{journal}{Phys. Rev. X}
  \textbf{\bibinfo{volume}{8}}, \bibinfo{pages}{021005} (\bibinfo{year}{2018}).

\bibitem[{\citenamefont{Goto}(2016{\natexlab{b}})}]{Goto2016a}
\bibinfo{author}{\bibfnamefont{H.}~\bibnamefont{Goto}}, \bibinfo{journal}{Phys.
  Rev. A} \textbf{\bibinfo{volume}{93}}, \bibinfo{pages}{050301}
  (\bibinfo{year}{2016}{\natexlab{b}}).

\bibitem[{\citenamefont{Dykman et~al.}(1998)\citenamefont{Dykman, Maloney,
  Smelyanskiy, and Silverstein}}]{Dykman1998a}
\bibinfo{author}{\bibfnamefont{M.~I.} \bibnamefont{Dykman}},
  \bibinfo{author}{\bibfnamefont{C.~M.} \bibnamefont{Maloney}},
  \bibinfo{author}{\bibfnamefont{V.~N.} \bibnamefont{Smelyanskiy}},
  \bibnamefont{and}
  \bibinfo{author}{\bibfnamefont{M.}~\bibnamefont{Silverstein}},
  \bibinfo{journal}{Phys. Rev. E} \textbf{\bibinfo{volume}{57}},
  \bibinfo{pages}{5202} (\bibinfo{year}{1998}).

\bibitem[{\citenamefont{Wustmann and Shumeiko}(2013)}]{Wustmann2013b}
\bibinfo{author}{\bibfnamefont{W.}~\bibnamefont{Wustmann}} \bibnamefont{and}
  \bibinfo{author}{\bibfnamefont{V.}~\bibnamefont{Shumeiko}},
  \bibinfo{journal}{Phys. Rev. B} \textbf{\bibinfo{volume}{87}},
  \bibinfo{pages}{184501} (\bibinfo{year}{2013}).

\bibitem[{\citenamefont{Siddiqi et~al.}(2005)\citenamefont{Siddiqi, Vijay,
  Pierre, Wilson, Frunzio, Metcalfe, Rigetti, Schoelkopf, Devoret, Vion
  et~al.}}]{Siddiqi2005}
\bibinfo{author}{\bibfnamefont{I.}~\bibnamefont{Siddiqi}},
  \bibinfo{author}{\bibfnamefont{R.}~\bibnamefont{Vijay}},
  \bibinfo{author}{\bibfnamefont{F.}~\bibnamefont{Pierre}},
  \bibinfo{author}{\bibfnamefont{C.~M.} \bibnamefont{Wilson}},
  \bibinfo{author}{\bibfnamefont{L.}~\bibnamefont{Frunzio}},
  \bibinfo{author}{\bibfnamefont{M.}~\bibnamefont{Metcalfe}},
  \bibinfo{author}{\bibfnamefont{C.}~\bibnamefont{Rigetti}},
  \bibinfo{author}{\bibfnamefont{R.~J.} \bibnamefont{Schoelkopf}},
  \bibinfo{author}{\bibfnamefont{M.~H.} \bibnamefont{Devoret}},
  \bibinfo{author}{\bibfnamefont{D.}~\bibnamefont{Vion}}, \bibnamefont{et~al.},
  \bibinfo{journal}{Phys. Rev. Lett.} \textbf{\bibinfo{volume}{94}},
  \bibinfo{pages}{027005} (\bibinfo{year}{2005}).

\bibitem[{\citenamefont{Wilson et~al.}(2010)\citenamefont{Wilson, Duty,
  Sandberg, Persson, Shumeiko, and Delsing}}]{Wilson2010}
\bibinfo{author}{\bibfnamefont{C.~M.} \bibnamefont{Wilson}},
  \bibinfo{author}{\bibfnamefont{T.}~\bibnamefont{Duty}},
  \bibinfo{author}{\bibfnamefont{M.}~\bibnamefont{Sandberg}},
  \bibinfo{author}{\bibfnamefont{F.}~\bibnamefont{Persson}},
  \bibinfo{author}{\bibfnamefont{V.}~\bibnamefont{Shumeiko}}, \bibnamefont{and}
  \bibinfo{author}{\bibfnamefont{P.}~\bibnamefont{Delsing}},
  \bibinfo{journal}{Phys. Rev. Lett.} \textbf{\bibinfo{volume}{105}},
  \bibinfo{pages}{233907} (\bibinfo{year}{2010}).

\bibitem[{\citenamefont{Zhang and Dykman}(2017)}]{Zhang2017}
\bibinfo{author}{\bibfnamefont{Y.}~\bibnamefont{Zhang}} \bibnamefont{and}
  \bibinfo{author}{\bibfnamefont{M.~I.} \bibnamefont{Dykman}},
  \bibinfo{journal}{Phys. Rev. A} \textbf{\bibinfo{volume}{95}},
  \bibinfo{pages}{053841} (\bibinfo{year}{2017}).

\bibitem[{\citenamefont{Wang et~al.}(2019)\citenamefont{Wang, Pechal, Wollack,
  Arrangoiz-Arriola, Gao, Lee, and Safavi-Naeini}}]{Wang2019a}
\bibinfo{author}{\bibfnamefont{Z.}~\bibnamefont{Wang}},
  \bibinfo{author}{\bibfnamefont{M.}~\bibnamefont{Pechal}},
  \bibinfo{author}{\bibfnamefont{E.~A.} \bibnamefont{Wollack}},
  \bibinfo{author}{\bibfnamefont{P.}~\bibnamefont{Arrangoiz-Arriola}},
  \bibinfo{author}{\bibfnamefont{M.}~\bibnamefont{Gao}},
  \bibinfo{author}{\bibfnamefont{N.~R.} \bibnamefont{Lee}}, \bibnamefont{and}
  \bibinfo{author}{\bibfnamefont{A.~H.} \bibnamefont{Safavi-Naeini}},
  \bibinfo{journal}{Phys. Rev. X} \textbf{\bibinfo{volume}{9}},
  \bibinfo{pages}{021049} (\bibinfo{year}{2019}).

\bibitem[{\citenamefont{Zurek}(2003)}]{Zurek2003}
\bibinfo{author}{\bibfnamefont{W.~H.} \bibnamefont{Zurek}},
  \bibinfo{journal}{Rev. Mod. Phys.} \textbf{\bibinfo{volume}{75}},
  \bibinfo{pages}{715} (\bibinfo{year}{2003}).

\bibitem[{\citenamefont{Munro et~al.}(2002)\citenamefont{Munro, Nemoto,
  Milburn, and Braunstein}}]{Munro2002a}
\bibinfo{author}{\bibfnamefont{W.~J.} \bibnamefont{Munro}},
  \bibinfo{author}{\bibfnamefont{K.}~\bibnamefont{Nemoto}},
  \bibinfo{author}{\bibfnamefont{G.~J.} \bibnamefont{Milburn}},
  \bibnamefont{and} \bibinfo{author}{\bibfnamefont{S.~L.}
  \bibnamefont{Braunstein}}, \bibinfo{journal}{Phys. Rev. A}
  \textbf{\bibinfo{volume}{66}}, \bibinfo{pages}{023819}
  (\bibinfo{year}{2002}).

\bibitem[{\citenamefont{Dykman et~al.}(2018)\citenamefont{Dykman, Bruder,
  L\"orch, and Zhang}}]{Dykman2018}
\bibinfo{author}{\bibfnamefont{M.~I.} \bibnamefont{Dykman}},
  \bibinfo{author}{\bibfnamefont{C.}~\bibnamefont{Bruder}},
  \bibinfo{author}{\bibfnamefont{N.}~\bibnamefont{L\"orch}}, \bibnamefont{and}
  \bibinfo{author}{\bibfnamefont{Y.}~\bibnamefont{Zhang}},
  \bibinfo{journal}{Phys. Rev. B} \textbf{\bibinfo{volume}{98}},
  \bibinfo{pages}{195444} (\bibinfo{year}{2018}).

\bibitem[{\citenamefont{Puri et~al.}(2017{\natexlab{b}})\citenamefont{Puri,
  Andersen, Grimsmo, and Blais}}]{Puri2017e}
\bibinfo{author}{\bibfnamefont{S.}~\bibnamefont{Puri}},
  \bibinfo{author}{\bibfnamefont{C.~K.} \bibnamefont{Andersen}},
  \bibinfo{author}{\bibfnamefont{A.~L.} \bibnamefont{Grimsmo}},
  \bibnamefont{and} \bibinfo{author}{\bibfnamefont{A.}~\bibnamefont{Blais}},
  \bibinfo{journal}{Nat. Commun.} \textbf{\bibinfo{volume}{8}},
  \bibinfo{pages}{15785} (\bibinfo{year}{2017}{\natexlab{b}}).

\bibitem[{\citenamefont{Marandi et~al.}(2014)\citenamefont{Marandi, Wang,
  Takata, Byer, and Yamamoto}}]{Marandi2014}
\bibinfo{author}{\bibfnamefont{A.}~\bibnamefont{Marandi}},
  \bibinfo{author}{\bibfnamefont{Z.}~\bibnamefont{Wang}},
  \bibinfo{author}{\bibfnamefont{K.}~\bibnamefont{Takata}},
  \bibinfo{author}{\bibfnamefont{R.~L.} \bibnamefont{Byer}}, \bibnamefont{and}
  \bibinfo{author}{\bibfnamefont{Y.}~\bibnamefont{Yamamoto}},
  \bibinfo{journal}{Nat. Photonics} \textbf{\bibinfo{volume}{8}},
  \bibinfo{pages}{937} (\bibinfo{year}{2014}).

\bibitem[{\citenamefont{Nielsen and Chuang}(2000)}]{Chuang2000}
\bibinfo{author}{\bibfnamefont{M.}~\bibnamefont{Nielsen}} \bibnamefont{and}
  \bibinfo{author}{\bibfnamefont{I.}~\bibnamefont{Chuang}},
  \emph{\bibinfo{title}{Quantum Computation and Quantum Information}}
  (\bibinfo{publisher}{{Cambridge University Press}}, \bibinfo{year}{2000}).

\bibitem[{\citenamefont{Frattini et~al.}(2018)\citenamefont{Frattini, Sivak,
  Lingenfelter, Shankar, and Devoret}}]{Frattini2018a}
\bibinfo{author}{\bibfnamefont{N.~E.} \bibnamefont{Frattini}},
  \bibinfo{author}{\bibfnamefont{V.~V.} \bibnamefont{Sivak}},
  \bibinfo{author}{\bibfnamefont{A.}~\bibnamefont{Lingenfelter}},
  \bibinfo{author}{\bibfnamefont{S.}~\bibnamefont{Shankar}}, \bibnamefont{and}
  \bibinfo{author}{\bibfnamefont{M.~H.} \bibnamefont{Devoret}},
  \bibinfo{journal}{Phys. Rev. Applied} \textbf{\bibinfo{volume}{10}},
  \bibinfo{pages}{054020} (\bibinfo{year}{2018}).

\bibitem[{\citenamefont{Leghtas et~al.}(2015)\citenamefont{Leghtas, Touzard,
  Pop, Kou, Vlastakis, Petrenko, Sliwa, Narla, Shankar, Hatridge
  et~al.}}]{Leghtas2015}
\bibinfo{author}{\bibfnamefont{Z.}~\bibnamefont{Leghtas}},
  \bibinfo{author}{\bibfnamefont{S.}~\bibnamefont{Touzard}},
  \bibinfo{author}{\bibfnamefont{I.~M.} \bibnamefont{Pop}},
  \bibinfo{author}{\bibfnamefont{A.}~\bibnamefont{Kou}},
  \bibinfo{author}{\bibfnamefont{B.}~\bibnamefont{Vlastakis}},
  \bibinfo{author}{\bibfnamefont{A.}~\bibnamefont{Petrenko}},
  \bibinfo{author}{\bibfnamefont{K.~M.} \bibnamefont{Sliwa}},
  \bibinfo{author}{\bibfnamefont{A.}~\bibnamefont{Narla}},
  \bibinfo{author}{\bibfnamefont{S.}~\bibnamefont{Shankar}},
  \bibinfo{author}{\bibfnamefont{M.~J.} \bibnamefont{Hatridge}},
  \bibnamefont{et~al.}, \bibinfo{journal}{Science}
  \textbf{\bibinfo{volume}{347}}, \bibinfo{pages}{853} (\bibinfo{year}{2015}).

\bibitem[{\citenamefont{Nigg et~al.}(20)\citenamefont{Nigg, Paik, Vlastakis,
  Kirchmair, Shankar, Frunzio, Devoret, Schoelkopf, and Girvin}}]{Nigg2012a}
\bibinfo{author}{\bibfnamefont{S.~E.} \bibnamefont{Nigg}},
  \bibinfo{author}{\bibfnamefont{H.}~\bibnamefont{Paik}},
  \bibinfo{author}{\bibfnamefont{B.}~\bibnamefont{Vlastakis}},
  \bibinfo{author}{\bibfnamefont{G.}~\bibnamefont{Kirchmair}},
  \bibinfo{author}{\bibfnamefont{S.}~\bibnamefont{Shankar}},
  \bibinfo{author}{\bibfnamefont{L.}~\bibnamefont{Frunzio}},
  \bibinfo{author}{\bibfnamefont{M.~H.} \bibnamefont{Devoret}},
  \bibinfo{author}{\bibfnamefont{R.~J.} \bibnamefont{Schoelkopf}},
  \bibnamefont{and} \bibinfo{author}{\bibfnamefont{S.~M.}
  \bibnamefont{Girvin}}, \bibinfo{journal}{Phys. Rev. Lett.}
  \textbf{\bibinfo{volume}{108}} (\bibinfo{year}{20}).

\bibitem[{\citenamefont{Campagne-Ibarcq
  et~al.}(2018)\citenamefont{Campagne-Ibarcq, Zalys-Geller, Narla, Shankar,
  Reinhold, Burkhart, Axline, Pfaff, Frunzio, Schoelkopf
  et~al.}}]{CampagneIbarcq2018a}
\bibinfo{author}{\bibfnamefont{P.}~\bibnamefont{Campagne-Ibarcq}},
  \bibinfo{author}{\bibfnamefont{E.}~\bibnamefont{Zalys-Geller}},
  \bibinfo{author}{\bibfnamefont{A.}~\bibnamefont{Narla}},
  \bibinfo{author}{\bibfnamefont{S.}~\bibnamefont{Shankar}},
  \bibinfo{author}{\bibfnamefont{P.}~\bibnamefont{Reinhold}},
  \bibinfo{author}{\bibfnamefont{L.}~\bibnamefont{Burkhart}},
  \bibinfo{author}{\bibfnamefont{C.}~\bibnamefont{Axline}},
  \bibinfo{author}{\bibfnamefont{W.}~\bibnamefont{Pfaff}},
  \bibinfo{author}{\bibfnamefont{L.}~\bibnamefont{Frunzio}},
  \bibinfo{author}{\bibfnamefont{R.}~\bibnamefont{Schoelkopf}},
  \bibnamefont{et~al.}, \bibinfo{journal}{Phys. Rev. Lett.}
  \textbf{\bibinfo{volume}{120}}, \bibinfo{pages}{200501}
  (\bibinfo{year}{2018}).

\bibitem[{\citenamefont{Geerlings et~al.}(2013)\citenamefont{Geerlings,
  Leghtas, Pop, Shankar, Frunzio, Schoelkopf, Mirrahimi, and
  Devoret}}]{Geerlings2013a}
\bibinfo{author}{\bibfnamefont{K.}~\bibnamefont{Geerlings}},
  \bibinfo{author}{\bibfnamefont{Z.}~\bibnamefont{Leghtas}},
  \bibinfo{author}{\bibfnamefont{I.~M.} \bibnamefont{Pop}},
  \bibinfo{author}{\bibfnamefont{S.}~\bibnamefont{Shankar}},
  \bibinfo{author}{\bibfnamefont{L.}~\bibnamefont{Frunzio}},
  \bibinfo{author}{\bibfnamefont{R.~J.} \bibnamefont{Schoelkopf}},
  \bibinfo{author}{\bibfnamefont{M.}~\bibnamefont{Mirrahimi}},
  \bibnamefont{and} \bibinfo{author}{\bibfnamefont{M.~H.}
  \bibnamefont{Devoret}}, \bibinfo{journal}{Phys. Rev. Lett.}
  \textbf{\bibinfo{volume}{110}}, \bibinfo{pages}{120501}
  (\bibinfo{year}{2013}).

\bibitem[{\citenamefont{Pfaff et~al.}(2017)\citenamefont{Pfaff, Axline,
  Burkhart, Vool, Reinhold, Frunzio, Jiang, Devoret, and
  Schoelkopf}}]{pfaff2017}
\bibinfo{author}{\bibfnamefont{W.}~\bibnamefont{Pfaff}},
  \bibinfo{author}{\bibfnamefont{C.~J.} \bibnamefont{Axline}},
  \bibinfo{author}{\bibfnamefont{L.~D.} \bibnamefont{Burkhart}},
  \bibinfo{author}{\bibfnamefont{U.}~\bibnamefont{Vool}},
  \bibinfo{author}{\bibfnamefont{P.}~\bibnamefont{Reinhold}},
  \bibinfo{author}{\bibfnamefont{L.}~\bibnamefont{Frunzio}},
  \bibinfo{author}{\bibfnamefont{L.}~\bibnamefont{Jiang}},
  \bibinfo{author}{\bibfnamefont{M.~H.} \bibnamefont{Devoret}},
  \bibnamefont{and} \bibinfo{author}{\bibfnamefont{R.~J.}
  \bibnamefont{Schoelkopf}}, \bibinfo{journal}{Nature Physics}
  \textbf{\bibinfo{volume}{13}} (\bibinfo{year}{2017}).

\bibitem[{\citenamefont{Gambetta et~al.}(2007)\citenamefont{Gambetta, Braff,
  Wallraff, Girvin, and Schoelkopf}}]{gambetta2007}
\bibinfo{author}{\bibfnamefont{J.}~\bibnamefont{Gambetta}},
  \bibinfo{author}{\bibfnamefont{W.~A.} \bibnamefont{Braff}},
  \bibinfo{author}{\bibfnamefont{A.}~\bibnamefont{Wallraff}},
  \bibinfo{author}{\bibfnamefont{S.~M.} \bibnamefont{Girvin}},
  \bibnamefont{and} \bibinfo{author}{\bibfnamefont{R.~J.}
  \bibnamefont{Schoelkopf}}, \bibinfo{journal}{Phys. Rev. A}
  \textbf{\bibinfo{volume}{76}}, \bibinfo{pages}{012325}
  (\bibinfo{year}{2007}).

\bibitem[{\citenamefont{Touzard et~al.}(2019)\citenamefont{Touzard, Kou,
  Frattini, Sivak, Puri, Grimm, Frunzio, Shankar, and Devoret}}]{Touzard2019}
\bibinfo{author}{\bibfnamefont{S.}~\bibnamefont{Touzard}},
  \bibinfo{author}{\bibfnamefont{A.}~\bibnamefont{Kou}},
  \bibinfo{author}{\bibfnamefont{N.~E.} \bibnamefont{Frattini}},
  \bibinfo{author}{\bibfnamefont{V.~V.} \bibnamefont{Sivak}},
  \bibinfo{author}{\bibfnamefont{S.}~\bibnamefont{Puri}},
  \bibinfo{author}{\bibfnamefont{A.}~\bibnamefont{Grimm}},
  \bibinfo{author}{\bibfnamefont{L.}~\bibnamefont{Frunzio}},
  \bibinfo{author}{\bibfnamefont{S.}~\bibnamefont{Shankar}}, \bibnamefont{and}
  \bibinfo{author}{\bibfnamefont{M.~H.} \bibnamefont{Devoret}},
  \bibinfo{journal}{Phys. Rev. Lett.} \textbf{\bibinfo{volume}{122}},
  \bibinfo{pages}{080502} (\bibinfo{year}{2019}).

\bibitem[{\citenamefont{Yan et~al.}(2013)\citenamefont{Yan, Gustavsson,
  Bylander, Jin, Yoshihara, Cory, Nakamura, Orlando, and Oliver}}]{yan2013}
\bibinfo{author}{\bibfnamefont{F.}~\bibnamefont{Yan}},
  \bibinfo{author}{\bibfnamefont{S.}~\bibnamefont{Gustavsson}},
  \bibinfo{author}{\bibfnamefont{J.}~\bibnamefont{Bylander}},
  \bibinfo{author}{\bibfnamefont{X.}~\bibnamefont{Jin}},
  \bibinfo{author}{\bibfnamefont{F.}~\bibnamefont{Yoshihara}},
  \bibinfo{author}{\bibfnamefont{D.~G.} \bibnamefont{Cory}},
  \bibinfo{author}{\bibfnamefont{Y.}~\bibnamefont{Nakamura}},
  \bibinfo{author}{\bibfnamefont{T.~P.} \bibnamefont{Orlando}},
  \bibnamefont{and} \bibinfo{author}{\bibfnamefont{W.~D.}
  \bibnamefont{Oliver}}, \bibinfo{journal}{Nature Communications}
  \textbf{\bibinfo{volume}{4}}, \bibinfo{pages}{2337} (\bibinfo{year}{2013}).

\end{thebibliography}

\begin{ac}
AG designed and carried out initial experiments with help from ST and SOM and designed the final experiment with input from NEF, MHD, and SP. AG and NEF fabricated the sample, performed measurements and analyzed the data used in the manuscript. AG, NEF and MHD wrote the manuscript with input from all authors.
\end{ac}

\begin{ack}
We acknowledge the contributions of Luke Burkhart, Philippe Campagne-Ibarcq, Alec Eickbusch, Luigi Frunzio,
Philip Reinhold, Kyle Serniak, and Yaxing Zhang. Facilities use was supported by YINQE and the Yale SEAS cleanroom. This work was supported by ARO under grant No. W911NF-18-1-0212 and grant No. W911NF-16-1-0349  and NSF under grant No. DMR-1609326.
We also acknowledge support of the Yale Quantum Institute.
\end{ack}

\onecolumngrid
\newpage
\begin{center}
	\textsc{\Large{Supplementary Information}}
\end{center}

\section{General encoding}
\label{sec:bloch}
In this section, we define the general cat-qubit Bloch sphere following the conventions of Ref.~\cite{Puri2018a} in terms of the eigenstates of Hamiltonian $\hat{H}_{\mathrm{cat}}$~(\eqone) without the approximation that $|\left< +\alpha | -\alpha \right>| = e^{-2\bar{n}} \ll 1$ where $\bar{n} = |\alpha|^2$ as used in the main text of the paper. It has been shown \cite{Puri2017} that the two states $\catpm = \calN^{\pm}_{\alpha} \left( \ket{+\alpha} \pm \ket{-\alpha}  \right)$ are degenerate eigenstates of $\hat{H}_{\mathrm{cat}}$, where $\alpha = \sqrt{\epsilon_2/K}$ is the amplitude of the coherent states in the superposition and $\calN^{\pm}_{\alpha} = 1/\sqrt{2(1 \pm e^{-2\bar{n}})}$ is the normalization coefficient to account for $|\left< +\alpha | -\alpha \right>| \neq 0$. These expressions are valid for all $\epsilon_2/K$, and in the limit $\epsilon_2/K \to 0$ they become $\catp \to \ket{n=0}$ and $\catm \to \ket{n=1}$. This validates the adiabatic mapping between the Fock-qubit and cat-qubit Bloch spheres, which share a common definition $\ket{\pm Z} = \catpm$ for their respective values of $\alpha$.

The definitions relating the general encoding to the Bloch sphere of Fig.~{\figone}a in the limit $p = \calN^+_{\alpha} / \calN^-_{\alpha} \to 1$ are:
\begin{align}
\ket{\pm X} &= \left( \catp \pm \catm \right)/\sqrt{2} \to \xpm\\
\ket{\pm Y} &= \left( \catp \pm i \catm \right)/\sqrt{2} \to \left( \ket{+\alpha} \mp i \ket{-\alpha} \right)/\sqrt{2} = \ypm \\
\ket{\pm Z} &= \catpm \to \left( \ket{+\alpha} \pm \ket{-\alpha} \right)/\sqrt{2}
\end{align}
This limit, where the coherent states $\xpm$ are orthogonal, is reached exponentially fast with $\bar{n}=|\alpha|^2$. For the photon numbers used in this work, $\bar{n}=\nbarfigtwo$ (Fig.~\figtwo \Rabivsphase,\cutsRabivsphase,\wigners) and $\bar{n}=\nbarRest $ (Fig.~\figthree,\figfour) the associated values are $p \approx 0.988$ and $p \approx 0.994$. It is in this sense that $\ket{\pm X} \approx \xpm$ are macroscopically distinct states, which translates to a protection of the degeneracy of $\ket{\pm Y}$ and $\ket{\pm Z}$ against locally correlated noisy environments, or equivalently the suppression of phase flips.

This suppression also becomes evident by expressing the action of the photon-loss jump operator $\hat{a}$ on the states of the cat-qubit Bloch sphere:
\begin{align}
\hat{a}  &= \alpha \left[p^{-1} \ket{\mathcal{C}_{\alpha}^{+}}\bra{\mathcal{C}_{\alpha}^{-}} + p \ket{\mathcal{C}_{\alpha}^{-}}\bra{\mathcal{C}_{\alpha}^{+}} \right]\\
&= \alpha \left[\frac{p^{-1}+p}{2} \hat{\sigma}_x +  \frac{p^{-1}-p}{2} i\hat{\sigma}_y\right]
\end{align}
%
where ${\hat{\sigma}_x = \ket{\mathcal{C}_{\alpha}^{+}}\bra{\mathcal{C}_{\alpha}^{-}}+\ket{\mathcal{C}_{\alpha}^{-}}\bra{\mathcal{C}_{\alpha}^{+}}}$ and ${\hat{\sigma}_y =  -i(\ket{\mathcal{C}_{\alpha}^{+}}\bra{\mathcal{C}_{\alpha}^{-}} - \ket{\mathcal{C}_{\alpha}^{-}}\bra{\mathcal{C}_{\alpha}^{+}})}$ are the Pauli operators acting on the states defined above. For the photon numbers presented in this work the prefactors are $(p^{-1}+p)/2\approx 1$ and $(p^{-1}-p)/2\approx 0.01$ validating the approximations made in the analytical treatments in main text and in some sections of this supplementary information.  Throughout this work we use the general encoding for numerical simulations.

Finally, we point out that two conventions for the orientation of the cat-qubit encoding on the Bloch-sphere are used in the literature. One is described above and used in this work. The other convention corresponds to the application of a ninety-degree rotation (or a Hadamard gate) to the basis states of our encoding such that the coherent states are along the Z-axis and the even and odd cat states $\catpm$ are along the X-axis of the new encoded qubit Bloch sphere~\cite{Guillaud2019,Lescanne2019}. In both conventions bit-flips are defined as stochastic $\pi$-rotations around the X-axis and and phase-flips as stochastic $\pi$-rotations around the Z-axis~\cite{Chuang2000}. Consequently, there is a difference in which of the two error channels is suppressed: In our encoding the phase-flip error channel is suppressed. Under the other convention it is the bit-flip error channel. This difference in convention is only a matter of convenience of nomenclature and does not change the error-protection capabilities of the cat-qubit.

\section{Full system Hamiltonian}
\label{sec:sysH}
Here we derive the effective system Hamiltonian from the combination of a nonlinear resonator (the Kerr-cat mode), a harmonic oscillator (the readout cavity) and several drives generating the parametric interactions used in this work. We start with the Hamiltonian
\begin{equation*}
\hat{H}_0 = \hat{H}_a + \hat{H}_b + \hat{H}\sub{c} + \hat{H}\sub{drives}.
\end{equation*}
The first term on the right-hand side of this equation is the Hamiltonian of the nonlinear resonator given by ${\hat{H}_a/\hbar = \omega_{a,0} \hat{a}^{\dag}_0 \hat{a}_0 + g_3(\hat{a}^{\dag}_0 + \hat{a}_0)^{3} + + g_4(\hat{a}^{\dag}_0 + \hat{a}_0)^{4}}$ with annihilation operator $\hat{a}_0$, frequency $\omega_{a,0}$, third-order nonlinearity $g_3$ and fourth-order nonlinearity $g_4$~\cite{Frattini2018a}. The second term ${\hat{H}_b/\hbar = \omega_{b,0} \hat{b}^{\dag}_0 \hat{b}_0}$ describes the readout cavity as a harmonic oscillator with frequency $\omega_{b,0}$ and annihilation operator $\hat{b}_0$. The third term ${\hat{H}\sub{c}/\hbar = g(\hat{a}^{\dag}_0 + \hat{a}_0)(\hat{b}^{\dag}_0 + \hat{b}_0)}$ gives the coupling of strength $g$ between the two modes. Finally, the last term ${\hat{H}\sub{drives}/\hbar = 2 \mathrm{Re}(e^{i\omega\sub{s} t})(\epsilon\sub{s} \hat{a}^{\dag}_0 + \epsilon^{*}\sub{s}\hat{a}_0) +2 \mathrm{Re}(e^{i\omega\sub{cr} t})(\epsilon\sub{cr} \hat{a}^{\dag}_0 + \epsilon^{*}\sub{cr}\hat{a}_0)}$ corresponds to two drives at respective frequencies $\omega\sub{s}$ and $\omega\sub{cr}$ and slowly varying complex envelopes $\epsilon\sub{s}$ and $\epsilon\sub{cr}$. The first drive will generate the effective single-mode squeezing used for stabilization and the second one will implement the effective resonant interaction used for cat-quadrature readout. Here we have, without loss of generality, assumed that the drives address the nonlinear resonator.

After bringing $\hat{H}\sub{c}$ into the Jaynes-Cummings form and performing the transformation ${\hat{a}_0=\hat{a}_1 + (g/\Delta)\hat{b}_1}$, ${\hat{b}_0=\hat{b}_1 - (g/\Delta)\hat{a}_1}$, where $\Delta=\omega_{b,0}-\omega_{a,0}$, the system Hamiltonian becomes
\begin{equation*}
\hat{H}_1/\hbar = \omega_a \hat{a}^{\dag}_1 \hat{a}_1 + \omega_b \hat{b}^{\dag}_1 \hat{b}_1 + g_3(\hat{f}_1 + \hat{f}^{\dag}_1)^{3} + g_4(\hat{f}_1 + \hat{f}^{\dag}_1)^{4} + 2 \mathrm{Re}(e^{i\omega\sub{s} t})(\epsilon\sub{s}\hat{f}^{\dag}_1 + \epsilon^{*}\sub{s}\hat{f}_1) +2 \mathrm{Re}(e^{i\omega\sub{cr} t})(\epsilon\sub{cr}\hat{f}^{\dag}_1 + \epsilon^{*}\sub{cr}\hat{f}_1),
\end{equation*}
with the new frequencies $\omega_a = \omega_{a,0}-2g^{2}/\Delta$ and $\omega_b = \omega_{b,0}+2g^{2}/\Delta$ as well as ${\hat{f}_1=\hat{a}_1 + (g/\Delta)\hat{b}_1}$. Since the two drives are independent and in particular the drive at frequency $\omega\sub{cr}$ is not always on, we separately perform two successive transformations to a displaced frame~\cite{Leghtas2015} for each mode. These transformations are given by ${\hat{a}_1=\hat{a}_2 + \xi_{a,\mathrm{s}}(t)}$, ${\hat{a}_2=\hat{a}_3 + \xi_{a,\mathrm{cr}}(t)}$, ${\hat{b}_1=\hat{b}_2 + \xi_{b,\mathrm{s}}(t)}$, and ${\hat{b}_2=\hat{b}_3 + \xi_{b,\mathrm{cr}}(t)}$, where the displacement amplitudes are chosen~\cite{Leghtas2015} such that the Hamiltonian takes on the more compact form
\begin{equation*}
\hat{H}\sub{disp}/\hbar = \omega_a \hat{a}^{\dag}_3 \hat{a}_3 + \omega_b \hat{b}^{\dag}_3 \hat{b}_3 + g_3\hat{f}^{3} + g_4\hat{f}^{4},
\end{equation*}
with ${\hat{f} = \hat{a}_3 + \xi_{a,\mathrm{s}}(t) + \xi_{a,\mathrm{cr}}(t) + (g/\Delta)(\hat{b}_3 + \xi_{b,\mathrm{s}}(t) + \xi_{b,\mathrm{cr}}(t)) +\mathrm{h.c.}} = \hat{a}_3 + (g/\Delta)\hat{b}_3 + \tilde{\xi}_{\mathrm{eff, s}}(t) + \tilde{\xi}_{\mathrm{eff, cr}}(t) +\mathrm{h.c.}$. For our system parameters, the sum and difference frequencies $|\omega_i\pm\omega_j|/2\pi$ for $i \in \{\mathrm{s,cr}\}$ and $j \in \{a,b\}$ are all larger or equal to $\approx\SI{3}{\giga \hertz}$, making them significantly larger than the loss rates of either mode $\kappa_a/2\pi\approx\SI{10}{\kilo \hertz}$ and $\kappa_b/2\pi\approx\SI{2}{\mega \hertz}$. Moreover $(g/\Delta)^2\approx 0.01 \ll 1$ (see section~\ref{sec:params}). Under these conditions the effective displacement amplitudes for $i \in \{\mathrm{s,cr}\}$ are
\begin{equation*}
\tilde{\xi}_{\mathrm{eff, i}}(t) \approx \left[\frac{\epsilon_i}{\omega_i-\omega_a} - \frac{\epsilon_i^{*}}{\omega_i+\omega_a} \right] e^{-i\omega_i t} = \xi_{\mathrm{eff, i}}e^{-i\omega_i t}, 
\end{equation*}
where $\epsilon_i$ and thus $\xi_{\mathrm{eff, i}}$ varies slowly with respect to $|\omega_i-\omega_a|$. In practice, with the exception of initial and final ramps and the $Z(\pi/2)$-gate, $\epsilon_i$ is constant in our experiment. Note that, as opposed to the case of parametric processes generated with four-wave mixing, we cannot neglect the second term in the above expression as the tone applied to generate the parametric processes is far detuned from the mode frequency.

We now go into the rotating frames defined by $\hat{a}_3=\hat{a}e^{-i\frac{\omega\sub{s}}{2}t}$ and $\hat{b}_3=\hat{b}e^{-i\omega_bt}$, expand the the nonlinear terms and perform the rotating wave approximation for $\frac{\omega\sub{s}}{2}\approx \omega_a$ and $\omega\sub{cr} = \omega_b - \frac{\omega\sub{s}}{2}$. This yields the full system Hamiltonian,
%
%
\begin{equation}
\hat{H}/\hbar = \Delta_{as}\hat{a}^{\dag} \hat{a} -K \hat{a}^{\dag 2} \hat{a}^{2} +  \epsilon_{2}\hat{a}^{\dag 2} + \epsilon_{2}^{*}\hat{a}^{2}  - \chi_{ab}\hat{a}^{\dag}\hat{a}\hat{b}^{\dag}\hat{b}  -4K\hat{a}^{\dag} \hat{a} (|\xi_{\mathrm{eff, s}}|^{2}+|\xi_{\mathrm{eff, cr}}|^{2}) + g\sub{cr}\hat{a}^{\dag} \hat{b} + g\sub{cr}^{*}\hat{a} \hat{b}^{\dag}
\label{eq:Hfull}
\end{equation}
In the above expression, the first term represents a detuning between half the frequency of the tone generating the effective squeezing drive and the undriven mode frequency given by $\Delta_{as}=\omega_a-\frac{\omega_s}{2}$, the second term is the Kerr nonlinearity with $K=-6g_4$, and the third and fourth terms are the squeezing drive of strength $\epsilon_2 = 3g_3 \xi_{\mathrm{eff, s}}$. The fifth term is the cross-Kerr interaction of strength $\chi_{ab}=-24g_4(\frac{g}{\Delta})^{2}$ between the nonlinear resonator and the readout cavity used for dispersive readout of the Fock qubit. The next two terms are Stark shifts of the nonlinear resonator due to the two applied tones. The last two terms are the generated effective resonant interaction  $g\sub{cr}=6g_3\frac{g}{\Delta}\xi_{\mathrm{eff, cr}}^{*}$ between the two modes used for cat-quadrature readout. Here we have assumed that the Stark shifts of the readout cavity as well as its induced self-Kerr are negligible, which is confirmed by our experiments.

When no cat-quadrature readout drive is applied ($\xi_{\mathrm{eff, cr}}=0$) and the readout cavity is in the vacuum state (which is the case in our experiment whenever no readout is performed) equation~(\ref{eq:Hfull}) simplifies to
\begin{equation}
\hat{H}\sub{s}/\hbar = \Delta_{as}\hat{a}^{\dag} \hat{a} -K \hat{a}^{\dag 2} \hat{a}^{2} +  \epsilon_{2}\hat{a}^{\dag 2} + \epsilon_{2}^{*}\hat{a}^{2} -4K\hat{a}^{\dag} \hat{a}|\xi_{\mathrm{eff, s}}|^{2}.
\label{eq:Heff}
\end{equation}
This implements the Hamiltonian~({\eqone}) from the main text when we chose the frequency detuning $\Delta_{as}$ such that it compensates for the Stark shift. In the experiment, this detuning can take on a more complicated form, because the pump-dependent frequency shift acquires additional contributions due to the single mode squeezing drive. Additionally, the Kerr-nonlinearty also acquires a perturbative second-order correction from the third-order nonlinearity~\cite{Frattini2018a}. For the data presented in Fig.~{\figthree} and Fig.~{\figfour} of the main text we perform a tuneup experiment (see section~\ref{sec:tuneup}) to eliminate all effective detunings. For the achieved squeezing-drive strength of $\epsilon_2=\epstworest$ we have to detune our drive by $\Delta_{as}\approx\detuning$.

\section{Energy gap and spectrum}
\label{sec:gap}
In this section we give an estimate of the energy gap separating the states $\{\catp, \catm \}$ from the rest of the energy spectrum of the ideal Hamiltonian~({\eqone}) and compare the spectrum obtained from numerical diagonalization of the system Hamiltonian~(\ref{eq:Heff}) to experimental data.

To obtain an estimate of the gap we follow the derivation given in reference~\cite{Puri2018a}. Applying a displacement transformation $\hat{a} = \hat{a}' \pm \alpha$ to equation~({\eqone}) and using $\alpha = \sqrt{\epsilon_2/K}$ yields the Hamiltonian
\begin{equation}
\hat{H}'/\hbar = -4K\alpha^{2} \hat{a}'^{\dag} \hat{a}' \mp 2\alpha K (\hat{a}'^{\dag 2}\hat{a}' + \hat{a}'^{\dag}\hat{a}'^{2}) -K\hat{a}'^{\dag 2}\hat{a}'^{2}.
\label{eq:Hlimit}
\end{equation}
The vacuum state $\ket{0}$ is an eigenstate of this Hamiltonian. In the non-displaced frame this corresponds to the states $\xpm=\mathit{D}(\pm \alpha)\ket{0}$, where $\mathit{D}(\pm \alpha)$ is the displacement operator. In the limit of large $\alpha$ this becomes the Hamiltonian of an effective harmonic oscillator of frequency $-4K\alpha^{2}$. Then the next-closest eigenstate is the Fock state $\ket{1}$ which corresponds to a displaced state $\mathit{D}(\pm \alpha)\ket{1}$ in the original frame. The energy gap is the energy difference between these states given by $-4K\alpha^{2}=-4K\bar{n}$.

\begin{figure}[h]
	\includegraphics[angle = 0, width = \figwidthWide]{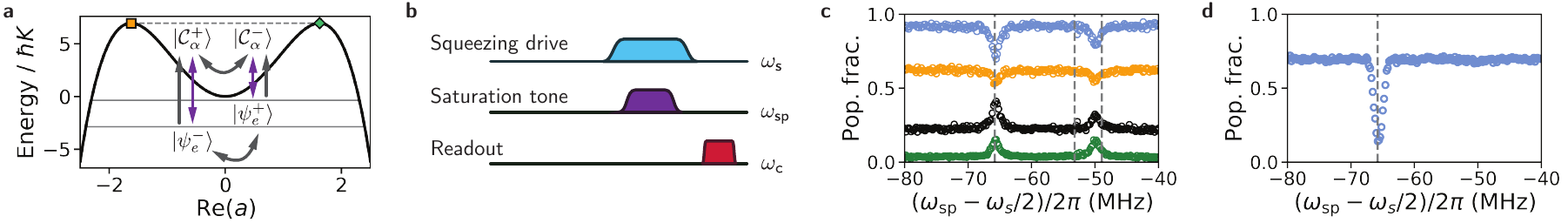}
	\caption{\label{SIfig:gap}  \textbf{Kerr-cat qubit gap spectroscopy. a},  Solid black line: Energy dependence of equation~(\ref{eq:Heff}) on the classical phase space coordinate $\mathrm{Re}(a)$ for $\mathrm{Im}(a)=0$. Symbols indicate the coherent state amplitudes $\pm \alpha$. Gray lines are the result of the numerical diagonalization of the Hamiltonian. The dashed gray line marks the degenerate states $\catpm$. The two solid lines indicate the energies of the two closest eigenstates. Purple arrows symbolize the transitions due to the single photon drive, while gray arrows symbolize transitions due to photon loss. \textbf{b}, Pulse sequences for gap spectroscopy. The cat-qubit is initialized in $\catp$. Then an saturation tone of varying frequency $\omega\sub{sp}$ is applied followed by mapping back onto the Fock qubit and dispersive readout. \textbf{c}, Measured Population fraction of the Fock states $\ket{0}$ (blue circles), $\ket{1}$ (orange circles), $\ket{2}$ (green circles), $\ket{n>2}$ (black circles) at the end of the experiment. The curves are offset for visibility. Gray dashed lines indicate the predicted transition frequencies (see text). The frequency axis is given with respect to the Stark-shifted mode frequency. \textbf{d}, Measured population fraction of the Fock state $\ket{0}$ when instead of a saturation tone a short pulse is applied.}
\end{figure}

In our experiment we are not in the limit discussed above as $\alpha \approx 1.6$. In order to determine the energy spectrum we need to consider the full Hamiltonian~(\ref{eq:Heff}) using our system parameters given in section~\ref{sec:params} (we use the value $\epsilon_2\approx\SI{17.75}{\mega \hertz}$). In Fig.~\ref{SIfig:gap}\SIFIGspec we plot its energy dependence on the classical phase space coordinate $\mathrm{Re}(a)$ for $\mathrm{Im}(a)=0$ as a solid black line. This corresponds to a cut through Fig.~{\figone}\phasespace of the main text. We numerically diagonalize the Hamiltonian to obtain the energies of the degenerate states $\catpm$ (dashed gray line in the figure) and of the closest two excited states $\ket{\psi_e^{\pm}}\approx(\mathrm{D}(\alpha)\pm\mathrm{D}(-\alpha))\ket{1}$ (solid gray lines). The former can also be obtained from direct calculation by factorizing the Hamiltonian~({\eqone})~\cite{Puri2017}. We find that all states have well defined photon number parity (indicated by the $\pm$ signs and that the energy differences between the cat-states and the states $\ket{\psi_e^{\pm}}$ are $\SI{-48.9}{\mega \hertz}\times h$ and $\SI{-65.8}{\mega \hertz}\times h$ respectively. Note that while the wells in the figure are inverted in this frame, all dissipative processes bring the system from the states $\ket{\psi_e^{\pm}}$ to the states $\catpm$~\cite{Puri2018a}.

We perform spectroscopy of the gap using the pulse sequence shown in Fig.~\ref{SIfig:gap}\SIFIGpulsespec. First, we initialize the system in $\catp$ by ramping on the squeezing drive. Then, we apply a saturation tone of varying frequency $\omega\sub{sp}$ and length $\SI{20}{\micro \second}$. Finally, we map the system back onto the Fock qubit and perform dispersive readout. The latter can distinguish between the $\ket{0}$, $\ket{1}$, $\ket{2}$ and $\ket{n>2}$ states of the Fock qubit. A coherent single photon drive does not conserve parity and thus we expect to drive the transitions $\catp \leftrightarrow \ket{\psi_e^{-}}$ and $\catm \leftrightarrow \ket{\psi_e^{+}}$ as indicated in Fig.~\ref{SIfig:gap}\SIFIGspec. At the same time, single photon loss also causes transitions between states of opposite parity as indicated in the figure by gray arrows. Since the mapping between the Fock-qubit and the cat-qubit Hamiltonian does conserve parity we expect the states $\ket{\psi_e^{\pm}}$ to map onto the $\ket{2}$ and $\ket{3}$ states of the Fock qubit. The result of this experiment is shown in Fig.~\ref{SIfig:gap}\SIFIGcatPspec. At several frequencies, we see a decrease in the population fractions of the $\ket{0}$ and $\ket{1}$ Fock states (blue and orange circles) and an increase in the population fractions of the $\ket{2}$ and $\ket{>2}$ Fock states (green and black circles). The two main peaks appear at $\SI{-65.8}{\mega \hertz}$ and $\SI{-49.7}{\mega \hertz}$ below the Stark-shifted mode frequency which are very close to the predicted values marked by dashed gray lines. An additional weak feature appears in both the $\ket{0}$ and $\ket{>2}$ curves at $\SI{-52.9}{\mega \hertz}$. We interpret this as a two-photon transition from $\catp$ to another even parity state. Our numerical simulation indeed predicts such a state at an energy difference of $\SI{-106.4}{\mega \hertz}\times h$ which corresponds to a two-photon transition frequency of $\SI{-53.2}{\mega \hertz}$ (dashed gray line in the figure). Finally, we confirm the parity-dependent selection rule of the driven transitions by applying only a short pulse instead of a saturation tone. The corresponding data (see Fig.~\ref{SIfig:gap}\SIFIGcatMspec) shows a marked dip in the $\ket{0}$-state population only at the frequency predicted for the transition $\catp \leftrightarrow \ket{\psi_e^{-}}$, while as expected no response is observed for the transition $\catp \leftrightarrow \ket{\psi_e^{+}}$.

\section{Continuous X-rotation}
\subsection{Adiabaticity of the X-rotation}
To estimate the limit imposed on the continuous X-rotation we add a drive-term $\epsilon_x\hat{a}'^{\dag}+ \epsilon_x^{*}\hat{a}'$ to the displaced ideal Hamiltonian~(\ref{eq:Hlimit}). Using the same approximation as in section~\ref{sec:gap}, the evolution of $\hat{a}'$ is described by the quantum Langevin equation:
\begin{equation*}
\partial_t \hat{a}' = 4iK\alpha^{2} \hat{a}' -i\epsilon_x - \frac{\kappa_a}{2} \hat{a}' + \sqrt{\kappa_a}\hat{a}'\sub{in},  
\end{equation*}
where $\kappa_a / 2\pi \approx \SI{10}{\kilo \hertz}$ is the single-photon loss rate of the nonlinear resonator and $\hat{a}'\sub{in}$ is the standard delta-correlated Gaussian vacuum noise with $\langle\hat{a}'\sub{in}\rangle =0$. In the steady state, this translates to a displacement of the vacuum state by $\alpha' =\langle \hat{a}' \rangle = \frac{\epsilon_x}{4K\alpha^{2}+i\kappa_a/2}$. In order for the gate to be adiabatic, such a displacement should not happen meaning that $\pm \alpha$ are still the eigenstates of the Hamiltonian~({\eqone}) in the original frame. This establishes a condition $\epsilon_x \ll 4 K \alpha^{2}$ for gate adiabaticity. In the work at hand, the maximum drive strength is approximately one order of magnitude weaker than the smaller of two gaps described in section~\ref{sec:gap}. Note also that the steady state is only reached after a ring-up time of $1/\kappa_a=T_1\approx \bareTone$. The gate times presented in this work (on the order of $\approx\SI{10}{\nano \second}$) are significantly shorter thus further relaxing the adiabaticity constraint. It is under this adiabaticity condition, and in the limit $p \to 1$, that the Rabi rate can be calculated by considering the energy difference of the states $\xpm$ under the action of the $\epsilon_x$ drive-term (yielding equation~(\eqtwo) of the main text). For small values of $\alpha$ ($p<1$, see section~\ref{sec:bloch}), equation~(\eqtwo) is not a good estimate of the Rabi frequency as visible in Fig.~{\figtwo}\Rabivsamp.

\subsection{Numerical simulation}
\label{sec:num_sim}
Here we describe the numerical simulation performed to obtain the results shown in Fig.~{\figtwo}. As discussed in sections~\ref{sec:sysH} and~\ref{sec:tuneup}, we can compensate for all drive-related detunings by adjusting the drive frequencies. This is done for the data presented in Fig.~{\figthree} and Fig.~{\figfour}. However, when measuring the data shown in Fig.~{\figtwo} of the main text it would be impractical to perform this calibration for each drive strength $\epsilon_2$. Instead we apply the tone generating the squeezing drive at twice the unshifted mode frequency $\omega\sub{s}=2\omega_a$ and the X-rotation drive a frequency $\omega\sub{s}/2$. The resulting system Hamiltonian is 
\begin{equation*}
\hat{H}\sub{s}/\hbar = -K \hat{a}^{\dag 2} \hat{a}^{2} +  \epsilon_{2}\hat{a}^{\dag 2} + \epsilon_{2}^{*}\hat{a}^{2} -4K\hat{a}^{\dag} \hat{a}|\xi_{\mathrm{eff, s}}|^{2} + \epsilon_{x}\hat{a}^{\dag} + \epsilon_{x}^{*}\hat{a}.
\end{equation*}

The third term (detuning due to Stark shift) results in smaller photon numbers $\bar{n}$ than expected from equation~(\eqtwo) and we can thus not use the obtained Rabi frequency to directly calibrate $\epsilon_2$. Instead we fit a simulation of the above Hamiltonian including photon loss and gain (see section~\ref{sec:decoherence}) to the Rabi oscillations obtained for the maximum drive strength shown in Fig.~{\figtwo}\Rabivsamp with $\epsilon_2$ as a free parameter. Our simulation mimics the experiment using the same initial state of the Fock qubit ($\epop$ thermal $\ket{1}$-state population) and rise times for the tanh-ramps of the squeezing drive ($\tanhpump$) and the X-rotation drive ($\tanhrabi$) up to the point marked by a black arrow in Fig.~{\figtwo}\pulseRabi. At this point we compute $\bra{\mathcal{C}_{\alpha}^{+}}\rho\ket{\mathcal{C}_{\alpha}^{+}}$, where $\rho$ is the simulated density matrix of the non-linear resonator. In the experiment, the state $\ket{\mathcal{C}_{\alpha}^{+}}$ is mapped onto the state $\ket{0}$ of the Fock-qubit which is then measured. We scale the values obtained from simulation to match the readout contrast of the measured data.

\section{Characterization of gate operations}
\subsection{Process tomography}
We perform process tomography on the different operations demonstrated in this work by making use of the mapping between the Fock qubit and the cat qubit. As described in the main text, for each operation, we initialize the cat-qubit in the states corresponding to each of the six cardinal points of its Bloch sphere by first applying the corresponding pulses to the Fock qubit and then ramping on the squeezing drive. After the operation, we map the resulting state back onto the Fock qubit by ramping off the squeezing drive. We then apply one of three pulses (see Fig.~{\figthree}) to the Fock qubit. The phase of these pulses is calibrated to correctly align the states on the equator of the Kerr-cat qubit and the Fock qubit. This is followed by dispersive readout. The entire sequence inside the gray box shown in Fig.~{\figthree}a,c,e implements a measurement along each of the three axes of the Bloch sphere. Together with the six possible initial states this results in a total of 18 measured values.

In this process we make use of two properties of our system: First, the states $\catp$ and $\catm$ map back onto the $\ket{0}$- and $\ket{1}$-states of the Fock qubit, while other eigenstates of the Hamiltonian~(\ref{eq:Heff}) map onto higher-photon-number Fock states $\ket{n>1}$. Second, the dispersive readout can distinguish between these three cases in a single shot. We calculate all 18 expectation values by taking the difference between the average thresholded $\ket{0}$- and $\ket{1}$-state populations. We normalize the obtained values using the difference of the average thresholded $\ket{0}$- and $\ket{1}$-state measurement contrast of Rabi oscillations on the Fock qubit. The sum of the the average thresholded $\ket{0}$- and $\ket{1}$-state populations is used to calculate the total population which corresponds to the expectation value of the identity (see next section). This value is normalized by the total averaged thresholded $\ket{0}$- and $\ket{1}$-state population of the same Rabi oscillations on the Fock qubit. This approach discards any population fraction that has potentially leaked outside of the encoding space spanned by $\{\catp, \catm \}$ (e.g. to the states $\ket{\psi_e^{\pm}}$) and thus takes into account this type of infidelity. 

\subsection{Pauli-transfer-matrix method and fidelities}

We characterize the fidelities of the performed gates including state-preparation and measurement errors with the Pauli-transfer-matrix (PTM) method described in references~\cite{Chow2012a, Touzard2018a}. To this end, we first reconstruct the experimental density matrix for each cardinal point by computing
\begin{equation*}
\rho_i  = \frac{1}{2} (\langle I \rangle_i \mathds{1}+\langle X \rangle_i \sigma_x +\langle Y \rangle_i \sigma_y +\langle Z \rangle_i \sigma_z),
\end{equation*}
where $\langle X \rangle_i, \langle Y \rangle_i, \langle Z \rangle_i$ and ${\langle I \rangle_i}$ are the measured expectation values for a preparation on the cardinal point $i \in \{\pm X, \pm Y,\pm Z\}$ and $\mathds{1}$, $\sigma_x$, $\sigma_y$, $\sigma_z$ are the identity and Pauli matrices. We do not perform maximum likelihood estimation or any other normalization on the computed density matrices in order to not discard potential infidelities due to leakage out of the encoding space.

\begin{figure}[h]
	\includegraphics[angle = 0, width = \figwidthWide]{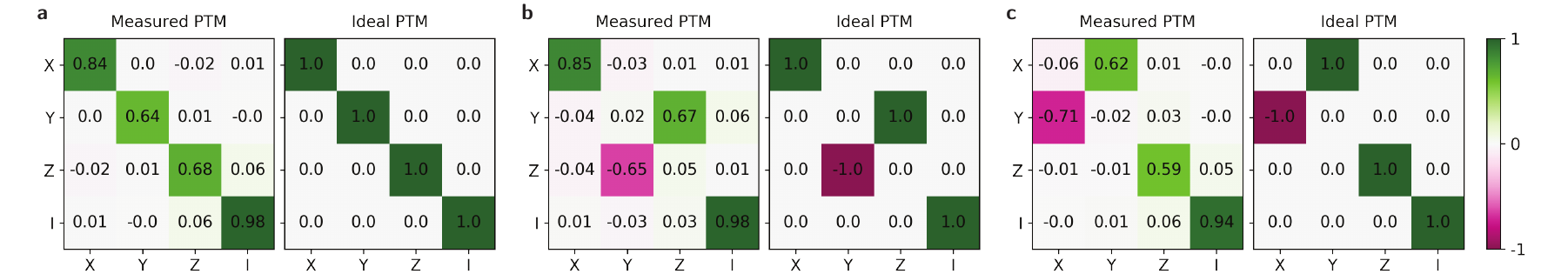}
	\caption{\label{SIfig:PTM}  \textbf{Pauli transfer matrix. a},\textbf{b},\textbf{c}, Measured ($\mathcal{R}^{\mathrm{exp}}$) and ideal ($\mathcal{R}^{\mathrm{ideal}}$) Pauli-transfer-matrix (PTM) representations of the mapping operation, the $X(\pi/2)$ rotation, and the $Z(\pi/2)$ rotation respectively. The three datasets correspond to the Bloch-sphere representations shown in Fig.~{\figthree}\blochMap, \blochX, and \blochZ of the main text.}
\end{figure}

Next, we express the experimental Pauli matrices (after application of the operation we wish to characterize) from the obtained density matrices as follows:
\begin{align*}
\mathds{1}\sub{exp}  &= \sum_i\rho_i/3\\
\sigma_{X\mathrm{,exp}}  &= \rho_{+X}-\rho_{-X}\\
\sigma_{Y\mathrm{,exp}}  &= \rho_{+Y}-\rho_{-Y}\\
\sigma_{Z\mathrm{,exp}}  &= \rho_{+Z}-\rho_{-Z}
\end{align*}
We can then calculate the $4\times4$ Pauli transfer matrix fully characterizing of the operation with components
\begin{equation*}
\mathcal{R}^{\mathrm{exp}}_{jk} = \frac{1}{2}\mathrm{Tr}(P_j^{\mathrm{init.}} P_k^{\mathrm{fin.}}),
\end{equation*}
where $P_j^{\mathrm{init.}}$ and $P_k^{\mathrm{fin.}}$ are the experimental Pauli matrices before and after the operation. We use the Pauli vector obtained for the mapping as  $P^{\mathrm{init.}}$. Finally, the fidelity $\mathcal{F}=\frac{1}{3}(\frac{1}{2}\mathrm{Tr}(\mathcal{R}^{\mathrm{ideal,T}}\mathcal{R}^{\mathrm{exp}})+1)$ is calculated by comparing the obtained experimental Pauli transfer matrix $\mathcal{R}^{\mathrm{exp}}$ to the ideal one $\mathcal{R}^{\mathrm{ideal}}$. We estimate the statistical error in fidelity with a bootstrapping approach by resampling each expectation value $10^{4}$ times from a Gaussian distribution with a width given by the standard measurement error which is $<0.006$ for all measured expectation values. The experimental and ideal PTMs for the operations presented in the main text are shown in Fig.~\ref{SIfig:PTM}. The associated fidelities are summarized in table~\ref{tab:fid}.

\begin{table}[h!]
	\centering
	\begin{tabular}{c|c|c}
		Operation 
		& Fidelity ($\si{\percent}$)
		& Error $\pm 3\sigma$ ($\si{\percent}$)\\ \hline
		Mapping  & 85.5 & $\pm0.5$ \\ \hline
		$X(\pi/2)$  & 85.7 & $\pm0.4$ \\ \hline
		$Z(\pi/2)$  & 81.1 & $\pm0.4$ \\
	\end{tabular}  
	\caption{\textbf{Fidelities and statistical errors}, Fidelities are computed using the Pauli-transfer-matrix approach described in the text. The errors are estimated using a bootstrap method.}
	\label{tab:fid}
\end{table}

\subsection{Additional gate characterization}
\begin{figure}[h]
	\includegraphics[angle = 0, width = \figwidth]{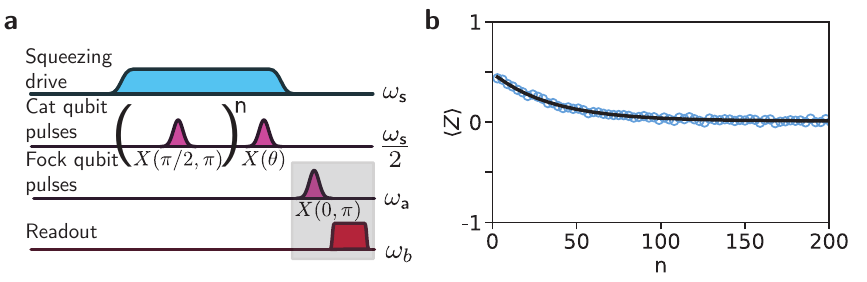}
	\caption{\label{SIfig:rb}  \textbf{Additional gate characterization a}, Pulse sequence to perform the following functions: (i) initialize the Kerr-cat-qubit in $\catp$, (ii) perform random sequence of $n$ gates (here either $X(\pi/2)$ or $X(\pi)$), (iii) perform an undo gate $X(\theta)$ where $\theta$ is calculated to map the state back to $\catp$, (iv) map onto the Fock qubit and perform symmetrized dispersive readout (grey box). \textbf{b}, Dependence of $\langle Z \rangle$ on the number of random gates applied $n$, where the average is an ensemble average over random sequences. Open blue circles are data and solid black line is a single exponential fit (see text).}
\end{figure}

The PTM estimate of the X-gate fidelity described above is mostly limited by state preparation and measurement errors as evidenced by the fact that its value is the same as the mapping fidelity (within error bars). To further investigate the quality of the $X(\theta)$ gate, we perform an additional characterization experiment. As depicted in the pulse sequence of Fig.~\ref{SIfig:rb}a, we initialize the cat-qubit in $\catp$, play a randomly generated sequence of $n$ gates uniformly selected from $\{ X(\pi/2), X(\pi) \}$, play an undo pulse that would set the cat-qubit back to $\catp$ if all gates were perfect, ramp back down to the Fock-qubit and perform a symmetrized dispersive readout. The dependence of the ensemble-averaged $\langle Z \rangle$ on the length of the random sequence $n$ is shown in Fig.~\ref{SIfig:rb}b. We fit the data with a single-exponential with decay constant $\tau_{n} = \tauRB$. The gate error for the $X(\theta)$ gate is then estimated via $r = (1-e^{-n/\tau_{n}})/2 = \rbError$, and the fidelity is $\mathcal{F}_X = 1-r = \rbFid$. The measured fidelity includes over-rotations, leakage out of the cat-encoding space and decoherence. Note that we estimate the probability for a  bitflip to occur during the gate operation to be $T_{X(\theta)}/\tau_{+Y} \approx 0.01$.

\section{Experimental setup}

\subsection{System details and sample design}
As described in the main text, our system includes a nonlinear resonator (see Fig.~{\figone}\snailmon), which hosts the quantum information, and a copper/aluminum readout cavity.  The former consists of a nonlinear inductance given by a superconducting asymmetric nonlinear element (SNAIL) which is shunted by a large capacitance coming from the two pads of dimension $\SI{2.2}{\milli \meter}\times\SI{0.6}{\milli \meter}$ visible in the photograph of Fig.~{\figone}\cavity. The SNAIL is a superconducting ring, placing a small Josephson junction (inductance $\SI{7.27}{\nano \henry}$) in parallel with three large junctions (inductance $\SI{0.8}{\nano \henry}$ per junction). The inductance values are estimated from room-temperature measurements of nominally identical test junctions on the same chip. This element has flux-tunable third- and fourth-order nonlinearities. We employ it here because it allows us to dissociate the strength of the single mode squeezing drive $\epsilon_2 = 3g_3 \xi_{\mathrm{eff, s}}$ from the Kerr-nonlinearity $K$ (see section~\ref{sec:sysH}). This makes it possible to stabilize cats with appreciable photon numbers $\bar{n}=\epsilon_2/K$ while limiting unwanted heating effects due to excessively strong drives amplitudes $\xi_{\mathrm{eff, s}}$~\cite{Sank2016a}. While the large capacitor pads are necessary to reduce the Kerr nonlinearity of the resonator, they lead to a large electric dipole moment. Strong coupling between the resonator and the readout cavity is not desirable for this experiment and so we avoid this by orienting the resonator perpendicularly to the electrical field direction of the lowest frequency cavity mode. In order to reintroduce a small cross-Kerr $\chi_{ab}$, which is necessary for dispersive readout, we offset the capacitor pads by $\delta=\SI{0.22}{\milli \meter}$. The next higher frequency mode has a field node at the position of the resonator and does not couple significantly to it. Higher frequency modes are far enough detuned to be negligible.

The readout cavity has two halves. Only the lower half is shown in Fig.~{\figone}\cavity. It is made from copper to let in the magnetic field used to tune the nonlinear resonator. The upper half is made from aluminum and is coupled to an aluminum WR-90 waveguide through an aperture. The waveguide itself couples to a transmission line through a $\SI{50}{\ohm}$-matched pin. It acts as a high-pass Purcell filter with a cutoff frequency of $\approx\SI{8.2}{\giga \hertz}$. This limits the decay of the nonlinear resonator into the continuum of modes of the transmission line, while letting pass the readout signal at frequency $\fres$ and the tone at $\approx\SI{12}{\giga \hertz}$ which generates the squeezing drive. The cavity additionally has a weakly coupled port realized with a standard coupling pin which allows us to drive the nonlinear resonator at $\approx\fqubit$ and apply the conversion drive at
$\approx\SI{2.9}{\giga \hertz}$ used for cat-quadrature readout. Both coupling-strengths are calibrated and tuned at room temperature using vector network analyzer transmission and reflection measurements. The entire system is shielded by both an aluminum- and a cryoperm-enclosure and thermalized to the base-stage of a dilution refrigerator at a temperature of $\approx\SI{18}{\milli \kelvin}$ as indicated in Fig.~\ref{SIfig:wiring}.

When designing the system we simulate its parameters in an iterative process. We first use a 3D electromagnetic field simulation software (ANSYS HFSS) and the black-box circuit quantization formalism~\cite{Nigg2012a} to predict the charging energy of the nonlinear resonator and its coupling strength to the cavity for the effective inductance of the SNAIL-element at a given flux point. Next, we calculate the resulting circuit parameters $g_3, K, \omega_a, \omega_b$ and the effective cross-Kerr $\chi_{ab}$ with a homemade program following the method described in reference~\cite{Frattini2017,Frattini2018a}. We then feed back onto the dimensions of the resonator and the inductances of the Josephson junctions until we achieve the desired parameters.

\subsection{Wiring diagram}
\begin{figure}[h]
	\includegraphics[angle = 0, width = \figwidthWide]{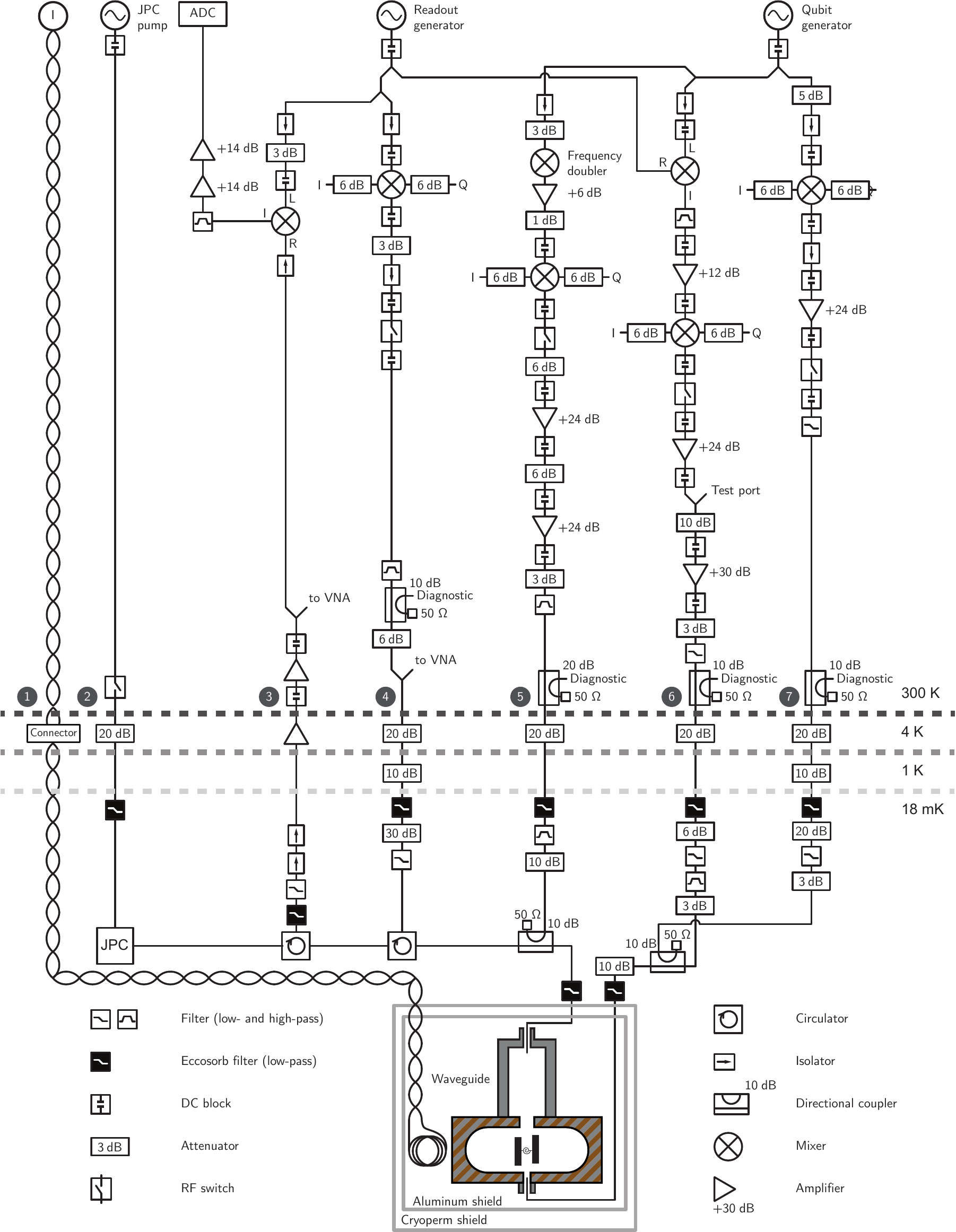}
	\caption{\label{SIfig:wiring}  \textbf{Wiring diagram}, A legend for the main elements is given. Attenuation or gain is indicated in $\mathrm{dB}$ where relevant. The different relevant temperature stages are indicated on the right-hand side. The sketch of the sample shows the nonlinear resonator inside the copper/aluminum readout cavity (brown/gray hashed) which is aperture-coupled to a waveguide. See text for description.}
\end{figure}

The wiring diagram shown in Fig.~{\ref{SIfig:wiring}} has seven branches which are numbered at the boundary with the dilution refrigerator in the schematic. From left to right in the figure they are: 1) the DC-current line used to generate an external magnetic field for the flux bias of the SNAIL, 2) the RF-line for the Josephson parametric converter (JPC) amplifier pump tone, 3) the readout output line, 4) the readout input line ($\omega_b$), 5) the line delivering the tone at frequency $\approx 2\omega_a$ which generates the squeezing drive, 6) the line delivering the tone at frequency $\omega\sub{cr}$ generating the interaction for cat-quadrature readout, and 7) the line used to resonantly address the nonlinear resonator at frequency $\approx \omega_a$.

The DC-line 1) connects a current source at room-temperature via a twisted-pair cable (normal metal from $\SI{300}{\kelvin}$ to $\SI{4}{\kelvin}$, superconducting below) to a superconducting magnet spool. This is used to thread a flux through the loop of the SNAIL element.

The next three lines 2) - 4) constitute a standard single-shot readout setup for a superconducting circuit experiment. A resonant tone (applied on line 4) is reflected of the readout cavity and acquires a qubit-state dependent phase shift. It is then routed to a JPC (which has a separate pump line numbered 2), reflected off with gain, routed to the output (line 3) and digitized after further amplification at $\SI{4}{\kelvin}$ and room temperature using an analog-to-digital converter (ADC). Two power splitters allow us to directly perform measurements with a vector network analyzer (VNA) for tuneup and characterization. When performing cat-quadrature readout, the same setup is used, but no drive is applied on line 4), because the readout signal is generated directly at the sample (see section~\ref{sec:cat_quad}).

The last three lines 5) - 7) deliver the essential drives of the experiment. These drives need to have high phase-stability with respect to each other and are thus created by mixing together the tones coming from two generators one of which (marked ``Readout generator'') is set to a frequency $\omega_b/2\pi+\SI{50}{\mega \hertz}$ while the other (marked ``Qubit generator'') is set to $\omega_a/2\pi+\SI{40}{\mega \hertz}$. Each branch going into the dilution refrigerator includes an IQ-mixer which takes its I and Q inputs from an FPGA-based DAC equipped with homemade software. These mixers modulate the the high-frequency tones with slow-varying envelopes ($\epsilon\sub{cr}$, $\epsilon\sub{s}$) and are calibrated such that they give out a single sideband at the respective desired frequency. The DAC has a sampling rate of $\approx \SI{500}{\mega \hertz}$ and is filtered with a low-pass filter at $\approx \SI{250}{\mega \hertz}$ to avoid aliasing. This limits the rise-time of the step-function used to perform the $Z(\pi/2)$-gate shown in Fig.~{\figtwo}\blochZ to $\approx \SI{4}{\nano \second}$ and allows us to create sidebands at a frequency difference of $\approx \pm \SI{125}{\mega \hertz}$ from the carrier tone.

We now describe these three lines in turn: Line 5) uses an amplifying frequency doubler (shown as two elements in the figure) to generate a tone at $2\omega_a/2\pi +\SI{80}{\mega \hertz}$ which is then modulated and converted to the frequency $\omega_s/2\pi \approx 2\omega_a/2\pi$. We use isolators and attenuators in order to suppress reflections and leakage of reflected tones to other branches. The losses introduced by these elements as well as the mixers are counterbalanced by adding several amplifiers to the branch which are chosen such that the maximum amplitude of the mixer output leads to an output power on the last amplifier just below its $\SI{1}{\dB}$ compression point (although the power used in the final experiment presented here is $<\SI{10}{\percent}$ of this maximum power). A switch helps to suppress the carrier leakage of the IQ-mixer when no drive is applied and a bandpass filter at the input of the dilution refrigerator is chosen such that it strongly suppresses the noise generated by this chain at all relevant mode frequencies of the experiment. Note that, because of the third-order nonlinear process used here, the frequency of the drive is far off-resonant from $\omega_a$ and $\omega_b$ making this approach possible. Just before the line enters the dilution refrigerator (DR) a directional coupler reroutes a fraction of the signal to dedicated spectrum analyzer for diagnostic and mixer tuneup. Further attenuation and filtering inside the DR is followed by a directional coupler which we use here as a broadband combiner to merge this line with the readout tone before entering the waveguide.

Line 6) is laid out according to similar principles as line 5). The carrier is generated by mixing the frequencies of the readout generator and the qubit generator to give $\omega_b/2\pi -\omega_a/2\pi + \SI{10}{\mega \hertz}$ which is modulated by the IQ-mixer to give $\omega\sub{cr}/2\pi$. The resulting drive is sent into the weakly coupled port of the system. The filters on this line are either low-pass with a cutoff frequency between $\omega\sub{cr}/2\pi$ and $\omega_a/2\pi$ or band-pass centered around $\omega\sub{cr}/2\pi$. 

The last line to be described here is branch 7) which is used both to resonantly address the Fock qubit as well as the Kerr-cat qubit when necessary (e.g. for the X-rotation described in the main text) following again similar principles as before. The filters on this line are low-pass with a cutoff situated between the frequencies of the nonlinear resonator and the readout cavity.

\section{System parameters}
\label{sec:params}

\begin{table}[h!]
	\centering
	\begin{tabular}{l|c|l}
		\textbf{Parameter} 
		& \textbf{Value}
		& \textbf{Method of estimate or measurement}\\ \hlineB{2}
		Nonlinear resonator frequency $\omega_a/2\pi$   & \fqubit & Two-tone spectroscopy \\ \hline
		Readout cavity frequency $\omega_b/2\pi$  & \fres & Direct RF reflection msmt. \\ \hlineB{2}
		Nonlinear resonator single-photon decay time $T_1$   & \bareTone & Std. coherence msmt. \\ \hline
		Nonlinear resonator transverse relaxation time $T_2$   & \bareTphi & Std. Ramsey coherence msmt. \\ \hline
		Nonlinear resonator transverse relaxation time (echo) $T_{2e}$   & \bareTecho & Std. Ramsey echo coherence msmt. \\ \hline
		Readout cavity linewidth (output coupling) $\kappa_{b,c}/2\pi$   & \kappaout & Direct RF reflection msmt. \\ \hline
		Readout cavity linewidth (other losses) $\kappa_{b,l}/2\pi$   & \kappaother & Direct RF reflection msmt. \\ \hlineB{2}
		Kerr-nonlinearity $K/2\pi$ & \kerr & Saturation spectro. of $\ket{0}\leftrightarrow\ket{2}$ transition  \\ \hline
		Third-order nonlinearity $g_3/2\pi$ & $\approx\gthree$ & Design simulation, fit of tuneup  \\ \hline
		Cross-Kerr $\chi_{ab}/2\pi$ & $\approx \SI{200}{\kilo \hertz}-\SI{250}{\kilo \hertz}$ & Design simulation, and msmt. ind. deph.~\cite{CampagneIbarcq2018a} \\ \hlineB{2}
		Cat-Rabi-drive strength $\epsilon_x/2\pi$ (Fig.~{\figtwo})  & \epsx & From Fock-qubit Rabi frequency \\
		Cat-Rabi-drive strength $\epsilon_x/2\pi$ (Fig.~{\figthree}, Fig.~{\figfour}) & \epsxGate & Extrapolated from above  \\ \hline
		Squeezing-drive strength $\epsilon_2/2\pi$ (Fig.~{\figtwo})  & \epstwofigtwo & Fit to simulation (section~\ref{sec:num_sim}) \\
		Squeezing-drive strength $\epsilon_2/2\pi$ (Fig.~{\figthree}, Fig.~{\figfour}) & \SI{17.5}{\mega \hertz}-$\SI{18}{\mega \hertz}$ & From equation~(\eqtwo), extrapolated from above   \\ \hline
		Detuning to compensate Stark shifts $\Delta_{as}/2\pi$ & \detuning & Tuneup experiment (section~\ref{sec:tuneup})  \\ \hline
		Frequency-conversion interaction strength $g\sub{cr}/2\pi$ & \g & Fock-qubit decay under driving  (section~\ref{sec:cat_quad}) \\ \hline
	\end{tabular}  
	\caption{\textbf{System parameters.} Summary of the main system parameters discussed in the main text and section~\ref{sec:sysH} with a short summary of how they are measured or estimated. More details and additional parameters are given in the text. Bolder horizontal lines separate from top to bottom: system frequencies, coherence parameters, nonlinearities, and parameters of the system under driving.}
	\label{tab:params}
\end{table}

In this section we give the parameters of our system as summarized in Table~\ref{tab:params}. All parameters are given for the flux-bias used for all experiments presented in this work corresponding to a flux through the SNAIL loop of $\Phi \approx \phiSNAIL$, where $\Phi_0=h/2e$ is the superconducting magnetic flux quantum. We calibrate the flux-bias from a multi-period sweep of the small inherited flux-dependence of the readout-cavity frequency.

The coherence times of the nonlinear resonator (i.e. the Fock qubit) are measured using standard experiments. The values presented in the table were obtained during the same cooldown as the data presented in the rest of this work. Over several cooldowns we observe variations of $\approx \pm \SI{3}{\micro \second}$ on the single-photon decay time and $\approx \pm \SI{1}{\micro \second}$ on the transverse relaxation time for comparable flux-biases. In general, the latter time becomes shorter when tuning to flux-bias values where $\partial\omega_a/\partial \Phi$ increases indicating that it is limited by flux-noise.

We extract the total cavity linewidth at base temperature by performing a reflection measurement of the resonator response and fitting the resulting circle in the complex plane. This yields the part of the linewidth coming from the coupling to the waveguide $\kappa_{b,c}/2\pi=\kappaout$ and the part due to other losses $\kappa_{b,l}/2\pi=\kappaother$. From room-temperature calibration we estimate that the weakly coupled pin accounts for $\approx\kappapin$ of the latter number.

The Kerr-nonlinearity is measured by applying a strong saturation tone of varying frequency $\omega\sub{sat}$ to the nonlinear resonator and measuring its response with dispersive readout. At a detuning of $\omega_a - \omega\sub{sat} = K$ this tone excites the two-photon transition $\ket{0}\leftrightarrow\ket{2}$ which allows us to determine the value of $K$.

We estimate the value of the third-order nonlinearity from our design simulation to lie between $\SI{15}{\mega \hertz}$ and $\SI{30}{\mega \hertz}$. The large margin in this estimate comes from the fact that the linear inductance of the leads between the SNAIL element and the capacitive pads is only roughly estimated in the simulation. We estimate the experimental value of $g_3$ by reproducing the tuneup-experiment described in section~\ref{sec:tuneup} in simulation (using equation~(\ref{eq:Heff})). We fix all parameters including $\epsilon_2 = 3g_3 \xi_{\mathrm{eff, s}}$ and vary $g_3$ until the effective Stark shift matches the detuning $\Delta_{as}$ such that we observe no spurious Z-rotation. This procedure yields a value of $g_3/2\pi \approx \gthree$ and is sensitive to variations in this estimate on the order of $\approx \pm \SI{1}{\mega \hertz}$. This also lets us estimate the dimensionless pump strength used in the experiment as $\xi_{\mathrm{eff,s}}\approx0.29$.

The estimate of the cross-Kerr term based on our design simulation is $\chi_{ab}/2\pi \approx \SI{250}{\kilo \hertz}$. We verify this estimate by conducting a Ramsey experiment on the Fock qubit in the presence of a drive on the readout cavity. From the change in transverse relaxation time and oscillation frequency of the Ramsey curve we infer the effective cross-Kerr as described in reference~\cite{CampagneIbarcq2018a}. This experiment is performed for a single drive-power giving a rough estimate of the cross-Kerr which we round to the leading digit $\chi_{ab}\approx\SI{200}{\kilo \hertz}$. This also allows us to estimate $g/\Delta\approx0.1$ used in section~\ref{sec:sysH}.

The drive strength $\epsilon_x$ for the data presented in Fig.~{\figtwo} is measured directly from the Rabi-frequency of the Fock qubit 
$\Omega\sub{f}=2\epsilon_x$ when no squeezing drive is applied. The strength of the drive applied to perform the $X(\pi/2)$-rotation presented in Fig.~{\figthree} is obtained from a linear extrapolation of this value using the known mixer-amplitudes for both cases.

The value of $\epsilon_2/2\pi=\epstwofigtwo$ for the measurements presented in Fig.~{\figtwo} of the main text is extracted from simulation as described in section~\ref{sec:num_sim}. In this case, we cannot use the formula $\epsilon_2=K\bar{n}$ together with the photon-number calibration given by the Rabi-frequency of the cat qubit (see equation~(\eqtwo)), because the uncompensated Stark shifts reduce the photon number $\bar{n}=\nbarfigtwo$. In the case where we compensate for the Stark shifts (i.e. Fig.~{\figthree} and Fig.~{\figfour}), we can directly calculate the squeezing drive strength $\epsilon_2/2\pi\approx\SI{17.5}{\mega \hertz}$ from these expressions. Extrapolating the drive strength found for Fig.~{\figtwo}, yields a similar value of $\epsilon_2/2\pi\approx\SI{18}{\mega \hertz}$.

The detuning $\Delta_{as}/2\pi=\detuning$ is calibrated as described in section~\ref{sec:tuneup}. The coupling strength of the frequency-converting interaction used for cat-quadrature readout $g_{\mathrm{cr}}$ is measured as described in section~\ref{sec:cat_quad}. The remaining Hamiltonian parameter $\xi_{\mathrm{eff,cr}}\approx0.15$ is not directly measured but estimated using the expression for $g_{\mathrm{cr}}$.

The thermal photon number of the nonlinear resonator $n\sub{th}=\nth$ is determined with the Rabi population measurement method described in reference~\cite{Geerlings2013a}.

\section{Tuneup sequence}
\label{sec:tuneup}
In this section, we describe the key tuneup experiments employed to generate a cat-qubit with a specific photon number $\bar{n}=|\alpha|^2$ and calibrate the phases of the different RF drives to account for line dispersion and Stark shifts as seen in \eqref{eq:Heff}. The effect of Stark shifts can be described by considering the effect of a Hamiltonian term $\Delta \hat{a}^\dagger \hat{a}$ where $\Delta = \Delta_{as} - 4K |\xi_{\mathrm{eff, s}}|^{2}$ and $\Delta_{as} = \omega_a - \omega_s/2$. Projected in the cat-qubit basis, this Hamiltonian reduces to $\Omega_z \hat{\sigma}_z$ where the Rabi rate around the Z-axis is $\Omega_z = -4\Delta|\alpha|^2 e^{-2|\alpha|^2}$ \cite{Puri2017}. We use this effect in experiment to tune $\omega_s$ to minimize $\Omega_z$ for a given drive strength.

\begin{figure}[h]
	\includegraphics[angle = 0, width = \figwidth]{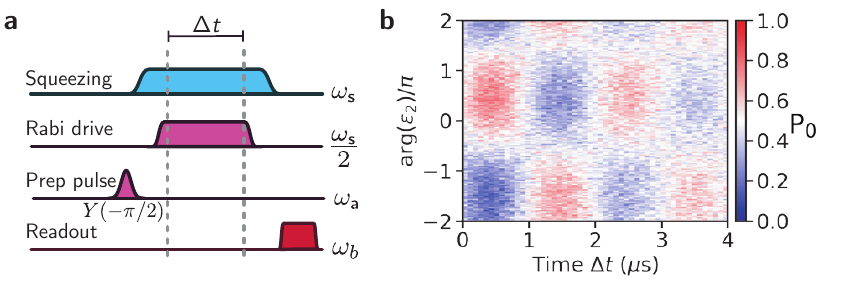}
	\caption{\label{SIfig:tuneup}  \textbf{Stabilization-drive phase tuneup a}, Pulse sequence to perform the following functions: (i) initialize the Kerr-cat-qubit on the equator of the Bloch sphere, (ii) drive Rabi oscillations for a varying time $\Delta t$, (iii) map onto the Fock qubit and perform dispersive readout. \textbf{b}, Dependence of Rabi oscillations on time $\Delta t$ and on the stabilization drive phase $\arg (\epsilon_2)$. The color scale gives the ground state population of the Fock qubit ($P_0$) at the end of the sequence.}
\end{figure}

The work flow is as follows: We choose a given frequency $\omega_s$ and drive strength for the drive used to generate the squeezing. We perform an experiment identical to the phase-dependant time-Rabi experiment in Fig.~{\figtwo}\Rabivsphase to calibrate the phase of $\epsilon_x$, which depends on the relative line dispersion between the qubit-drive and squeezing-drive lines, and measure $\Omega_x$ to extract $\bar{n}$ for these stabilization-drive parameters. Next, we calibrate the relative phase between the Fock-qubit pulses at $\omega_a$ and the cat-qubit at $\omega_s/2$ so that preparing $\ket{+X}$ on the Fock qubit and performing the adiabatic ramp up of $\epsilon_2$ maps to $\ket{+X}$ of the cat qubit. The pulse sequence used for this calibration is given in Fig.~\ref{SIfig:tuneup}a: prepare $\ket{+X}$ on the Fock qubit, adiatbatically ramp to the cat-qubit, apply Rabi drive $\epsilon_x$ with the previously calibrated phase for a varying time $\Delta t$, map back onto the Fock qubit and perform dispersive readout. We sweep the squeezing-drive phase $\arg{(\epsilon_2)}$ and the Rabi phase $\arg{(\epsilon_x)}$ together, maintaining the calibrated difference between them, and plot the resulting dependence of the Rabi oscillations in Fig.~\ref{SIfig:tuneup}b. The phase for which Rabi oscillations are suppressed (white horizontal lines) marks the phase for which the cat-qubit state was $\ket{+X}=\ket{\alpha}$, thus calibrating our mapping between the Fock and cat qubits. Note that the data are $4\pi$ periodic in the squeezing-drive phase $\arg \epsilon_2$, which is a result of the period-doubling phenomenon since there are two valid values of $\alpha$ in the frame rotating at $\omega_s/2$. For poor combinations of stabilization drive amplitude and frequency, the white lines in the data would be tilted due the the spurious Z-rotation at rate $\Omega_z$. This tuneup procedure can be iterated to minimize the measured $\Omega_z$ by choosing a new squeezing drive frequency or new $\epsilon_2$. The end result is a set of squeezing- and Rabi-drive parameters for a given cat-qubit of size $\bar{n}$ and the ability to prepare any point on the cat-qubit Bloch sphere through preparation of the Fock qubit and adiabatic mapping.

Finally, we discuss how to tune up the gates. For the $X(\theta)$ gate, a standard amplitude-Rabi experiment is performed where the qubit is prepared in $\catp$, a Gaussian pulse at $\omega_s/2$ of variable amplitude with the calibrated phase is applied, and Rabi oscillations are observed through mapping back to the Fock qubit and performing dispersive readout. This easily calibrates the pulse amplitude for a desired $\theta$. For the $Z(\pi/2)$ gate, we turn the stabilization drive off for $T_{Z(\pi/2)} = \zgateT \approx \pi/2K$. The free Kerr evolution under $-K\hat{a}^{\dagger2} \hat{a}$ for $\pi/2K$ takes for example $\ket{-Y}=\ym \to \ket{\alpha e^{i\varphi}}$ with $\varphi = -\pi/2 + \Delta_{as} T_{Z(\pi/2)}$ where the first term comes from the free Hamiltonian and the second comes from the absence of Stark shift while the stabilization drive is off. As such, when switching the stabilization drive back on, we boost it in phase by $2\varphi$ and redefine the cat-qubit frame such that $\ket{+X} = \ket{\alpha e^{i\varphi}}$. All future $X(\theta)$ gate pulses and all future Fock qubit pulses are also boosted by phase $\varphi$. In experiment, we optimize $\varphi$ to maximize the $Z(\pi/2)$ gate fidelity, but the calibrated value almost identically aligns to the above calculation.

\section{Cat-quadrature readout}
\label{sec:cat_quad}
\subsection{Parametric process}
In this section, we derive how adding a drive at $\omega_{\rm{cr}} = \omega_b - \omega_s/2$, while the squeezing drive is on, implements the quantum non-demolition (QND) cat-quadrature readout. The parametric process generated by this drive displaces the readout cavity (annihilation operator $\hat{b}$) conditioned on the $\hat{\sigma}_x$ state of the qubit.
To see this, we derive the interaction Hamiltonian in the cat-qubit basis and solve the resulting dynamics of the readout cavity. Starting from equation~(\eqthree), we express $\hat{a}$ in the cat-qubit basis (see section~\ref{sec:bloch}) and write the Hamiltonian projected on the cat-qubit Bloch sphere as:
\begin{align}
\hat{H}_{\rm{cr}}/\hbar &= ig\sub{cr}\alpha \left( \frac{p+p^{-1}}{2} \right) (\hat{b}^\dagger - \hat{b}) \hat{\sigma}_x - g\sub{cr}\alpha \left( \frac{p-p^{-1}}{2} \right) (\hat{b}^\dagger + \hat{b}) \hat{\sigma}_y \\
&\approx ig\sub{cr}\alpha (\hat{b}^\dagger - \hat{b}) \hat{\sigma}_x
\end{align}
where $g$ is proportional to the drive amplitude at $\omega_{\rm{cr}}$, and in the second line we have used $p\approx 1$ for modest $\bar{n}$ (see section~\ref{sec:bloch}) or equivalently that $\ket{\pm X} \approx \ket{\pm \alpha}$. Using this Hamiltonian, we describe the dynamics of $\hat{b}$ with the quantum Langevin equation:
\begin{align}
\partial_t \hat{b} &= \frac{i}{\hbar} [ \hat{H}_{\rm{cr}}, \hat{b} ] - \frac{\kappa_b}{2} \hat{b} + \sqrt{\kappa_b} \hat{b}\sub{in} \\
&\approx g\sub{cr}\alpha \hat{\sigma}_x - \frac{\kappa_b}{2} \hat{b} + \sqrt{\kappa_b} \hat{b}\sub{in}
\end{align}
where $\kappa_b / 2\pi = \kappatot$ is the total linewidth of the readout cavity, and $\hat{b}\sub{in}$ is the standard delta-correlated input field with the property $\langle \hat{b}\sub{in} \rangle = 0$ since we apply no drives at $\omega_b$ for this readout. In a semi-classical treatment, we then solve for the coherent-state amplitude $\beta = \langle \hat{b} \rangle$ in response to turning on $\omega_{\rm{cr}}$ at time $t=0$ giving
\begin{equation}
\beta(t, \sigma_x) = \frac{2g\sub{cr}\alpha}{\kappa_b}\sigma_x \left[ 1 - e^{-\kappa_b t/2} \right]
\end{equation}
where $\sigma_x = \pm 1$ encodes the two possible measured qubit states. The steady state value is therefore $\ket{\beta}$  with $\beta = \pm 2\alpha g\sub{cr} /\kappa_b$. We can use this steady state $\beta$ to estimate the validity of our projection into the cat-qubit manifold. In a mean-field treatment, we take $\hat{b} \to \beta$ in~(\eqthree) resulting in a Hamiltonian that implements a single-photon drive on the cat mode of the form $-ig\sub{cr}\beta (\hat{a}^\dagger - \hat{a})$. Out-of-manifold leakage caused by this drive is suppressed so long as $|g\sub{cr}\beta| \ll E_{\rm{gap}}/\hbar$; the same condition for suppressed leakage in the $X(\theta)$ Rabi gate. Moreover, this drive has a $\pi/2$ phase shift with respect to the optimal drive phase $\arg(\epsilon_x)$ and therefore does not generate spurious in-manifold rotations.

\subsection{Strength of coupling}
\begin{figure}
	\includegraphics[angle = 0, width = \figwidth]{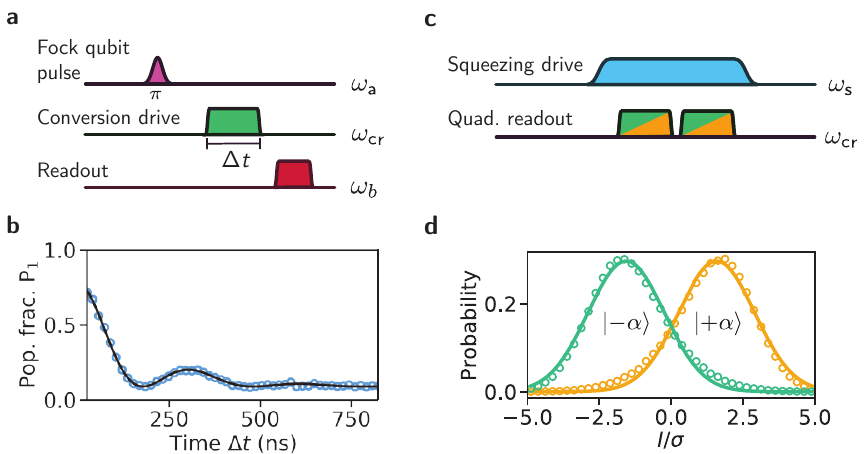}
	\caption{\label{SIfig:qswitch}  \textbf{Readout strength and QND-ness a}, Pulse sequence to measure the coupling strength $g\sub{cr}$. A $\pi$ pulse prepares $\ket{n=1}$, we apply the frequency-conversion drive at $\omega_{cr} = \omega_b - \omega_a$ for a variable time $\Delta t$, and perform dispersive readout.  \textbf{b}, Dependence of Fock-qubit $\ket{n=1}$ population ($P_1$) on $\Delta t$ for the pulse sequence in \textbf{a}. Open blue circles are measured data and the solid black line is a fit based on equation~(\ref{equ:g}).  \textbf{c}, Pulse sequence to test the QND-ness of the cat-quadrature readout: We prepare $\catp$, and perform two successive cat-quadrature readout pulses. \textbf{d}, Histogram of second readout postselected on the indicated result $\xpm$ of the first measurement. Open yellow (green) circles are measured data after finding $\xp$ ($\xm$) on the first measurement. Solid lines are Gaussian fits.}
\end{figure}

Here, we independently measure the induced coupling strength $g\sub{cr}$ in~(\eqthree). As can be seen in the pulse sequence of Fig.~\ref{SIfig:qswitch}a, we perform a $T_1$ measurement of the Fock qubit while turning the frequency-conversion drive on during the wait time. For $g\sub{cr} \ll \kappa_b$, the coupling to the readout cavity enhances the effective damping rate on the qubit and thus reduces the measured $T_{1\mathrm{eff}}$ to $1/T_{1\mathrm{eff}} = 1/T_1 + 4g\sub{cr}^2/\kappa_b$. This is often termed ``Q-switch'' and is commonly used for cooling. In our case, since $g\sub{cr} \approx \kappa_b$, we are actually driving Rabi oscillations between the bare Fock qubit and the readout cavity, which then leak out of the cavity at rate $\kappa_b$, see Fig.~\ref{SIfig:qswitch}b. Following the supplement of Ref.~\cite{pfaff2017}, we start from~(\eqthree) write the equations of motion for $\hat{a}$. We then solve the dynamics with the initial conditions of $\hat{a}(0)$ for the qubit and with the readout cavity in vacuum to get:
\begin{equation}
\hat{a}(t) = \frac{\hat{a}(0)}{\Omega} e^{-\kappa_b t/4} \left( \Omega \cosh{\frac{\Omega t}{4}}  + \kappa_b \sinh{\frac{\Omega t}{4}} \right)
\label{equ:g}
\end{equation}
where $\Omega = \sqrt{\kappa_b^2 - (4g\sub{cr})^2}$ is in general complex. Here we have assumed that the quadrature readout drive is on-resonance at $\omega_{cr} = \omega_b - \omega_a$. In practice, the Fock-qubit frequency is Stark-shifted by the quadrature readout drive (see Eq.~\ref{eq:Heff}), but we find the induced Stark shift $4K |\xi_{\mathrm{eff, cr}}|^{2} < \kappa_b$ to be small for our drive strength. In the experiment, we extract the Fock qubit population with dispersive readout, so we measure $\langle \hat{n}(t) \rangle = \langle \hat{a}^\dagger (t) \hat{a}(t) \rangle$. We fit the data with the corresponding expression resulting from equation~\ref{equ:g} with the initial condition $\langle \hat{n}(0) \rangle = 1$. Together with a scaling factor and an offset to compensate for readout contrast, we extract the only other free parameter $g\sub{cr}/2\pi = \g$. 

Importantly, as noted in the analysis in the previous section, despite $g \approx \kappa_b$, such Rabi oscillations do not occur when the squeezing drive is on and the cat-quadrature readout is performed. Intuitively, because $|g\sub{cr}\beta| \ll E\sub{gap}/\hbar$, the squeezing drive replenishes any photon that leaves the Kerr-cat mode before one Rabi-oscillation cycle can be completed. Further excitation of the Kerr-cat mode is then suppressed by the stabilization.

\subsection{Signal treatment}
Here, we briefly describe the signal treatment to generate histograms for the cat-quadrature readout and assign a binary value based on a threshold for single-shot measurements. These procedures are nearly identical to those performed for dispersive readout. To generate the histograms shown in Fig.~{\figfour}a, the full time trajectory of the readout signal associated with signal $\hat{b}\sub{out} = \sqrt{\kappa_b} \hat{b} - \hat{b}\sub{in}$ is recorded via heterodyne detection ($\SI{50}{\mega \hertz}$ intermediate frequency) and digitized for each cat-qubit preparation $\ket{\pm\alpha}$. We denote the average over many experimental shots of these two trajectories as $\beta\sub{out}^{\pm}(t)$. For each shot $m$, we assign a point on the IQ-plane $(I_m, Q_m)$ by integrating the measured signal over the full measurement time $\tau$ as: $\int_0^\tau \left[ \hat{b}\sub{out} e^{-i\varphi} +   \hat{b}^\dagger\sub{out} e^{i\varphi}  \right] K(t)dt$ where $\varphi= (0, \pi/2)$ for $(I_m, Q_m)$ and $K(t)$ is the integration envelope. We use the envelope $K(t) = \left( \beta\sub{out}^+ - \beta\sub{out}^- \right)^*$\cite{gambetta2007} to assign $(I_m,Q_m)$ and generate the histogram. We also offset the I and Q axes by a small independently measured amount coming from leakage inside our interferometer. This offset is however negligible compared to the width of the histograms and is only necessary to confirm that $\ket{\pm\alpha}$ indeed generate equal and opposite displacements on the readout cavity.

To calibrate a single-shot thresholded measurement, we integrate over the Q-axis resulting in the histograms shown in the bottom panel of Fig.~{\figfour}a. We fit each histogram to a Gaussian and scale the I-axis by the resulting width $\sigma$ (same for both $\ket{\pm\alpha}$). The mean between the two centers of the Gaussians is exactly $I=0$, where we then set our threshold for state assignment of future single-shot measurements.
\subsection{QND-ness}
An important property of a qubit readout is to be quantum non-demolition (QND). We test the QND-ness of the cat-quadrature readout by preparing $\catp$ and performing two cat-quadrature readout pulses one after the other as shown in the pulse sequence Fig.~\ref{SIfig:qswitch}c. The result of the first measurement generates histograms much like the bottom panel of Fig.~{\figfour}a. We postselect on the outcomes $\xpm$ for this first measurement using a stringent threshold. In Fig.~\ref{SIfig:qswitch}d, we plot the two histograms showing the outcomes of the second measurement contingent on this postselection. The I-axis is scaled by the $\sigma$ of the first measurement, and we find the second measurement has a 10\% larger $\sigma$ given by the Gaussian fit (solid). We calculate the QND-ness $\mathcal{Q} = (p(+\alpha | +\alpha) + p(-\alpha | -\alpha))/2= \qndness$, where $p(\pm \alpha | \pm \alpha)$ are the probabilities of the respective second measurement outcome conditioned on the first measurement~\cite{Touzard2019}. We note that the conversion process leads to a significantly decreased effective lifetime of the nonlinear resonator such that the phase flip rate due to single-photon loss becomes non-negligible in spite of the suppression factor $(p^{-1}-p)/2\approx 0.006$ for $\bar{n}=\nbarRest$ (see section~\ref{sec:bloch}). In order the quantify this effect, we simulate the master equation
\begin{equation*}
\dot{\hat{\rho}} = -\frac{i}{\hbar}[\hat{H}\sub{cat}+\hat{H}\sub{cr}, \hat{\rho}] +\kappa_b\mathcal{D}[\hat{b}]\hat{\rho},
\end{equation*}
where the two Hamiltonian terms correspond to equations~({\eqone}) and~(\eqthree) of the main text and $\mathcal{D}[\mathcal{O}]\hat{\rho}=\mathcal{O}\hat{\rho}\mathcal{O}^{\dag}-\frac{1}{2}\mathcal{O}^{\dag}\mathcal{O}\hat{\rho} -\frac{1}{2}\hat{\rho}\mathcal{O}^{\dag}\mathcal{O}$. Here we do not include decoherence of the nonlinear resonator. This yields a QND-ness of $\mathcal{Q} = 0.93$. If we add the additional decoherence terms present in equation~(\ref{eq:rho_full}) and use the parameters $n\sub{th}=0.08$ and $\kappa_{\phi,\mathrm{eff}}/2\pi=\SI{230}{\hertz}$ (see section~\ref{sec:decoherence}) we find $\mathcal{Q} = 0.90$. We expect an additional reduction of the measured QND-ness with respect to this value due to the finite separation of the histograms even in absence of additional phase-flips. These effects will be the topic of further study. A linear increase in the photon number exponentially suppresses these spurious effects~\cite{Puri2018a}, meaning that an increase in QND-ness (and also readout fidelity) should come with the mitigation of the drive-power-dependent heating (see section~\ref{sec:decoherence}) in future iterations of this system.

\section{Coherence measurements using dispersive readout}
\label{sec:disp}
In this section we measure the coherence properties presented in Fig.~{\figfour} of the main text in a complementary way by mapping the back onto the Fock qubit and performing dispersive readout. We additionally discuss leakage to states outside the cat qubit encoding.

\begin{figure}[h]
	\includegraphics[angle = 0, width = \figwidthWide]{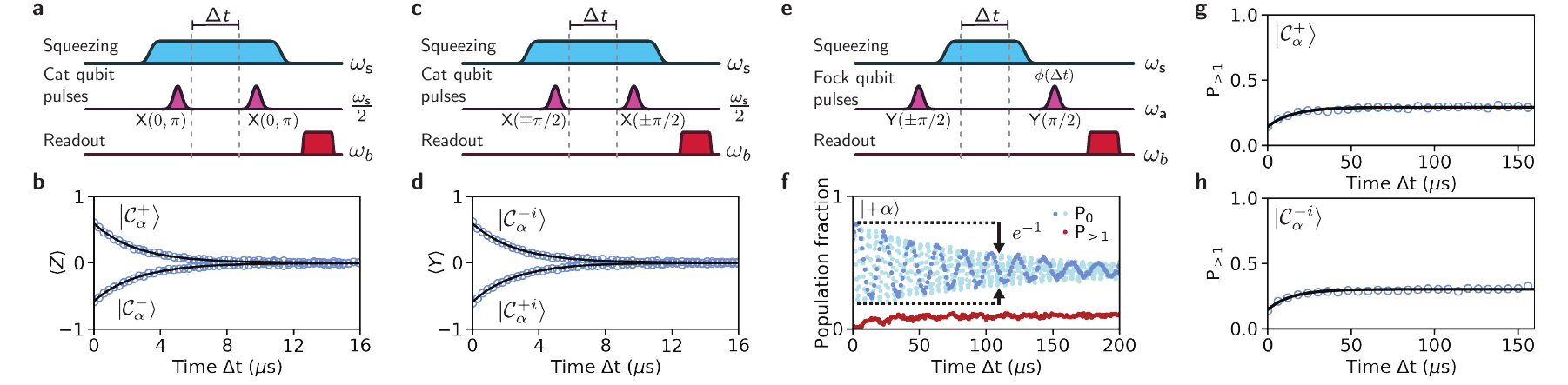}
	\caption{\label{SIfig:coh_disp}  \textbf{Coherence measurements using dispersive readout. a},\textbf{c}, Pulse sequences for measuring the lifetimes of the Schr\"{o}dinger-cat states along the Z- and Y-axis of the cat qubit. The first set of pulses (respectively $X(0),X(\pi),X(-\pi/2),X(\pi/2)$) initializes the cat qubit in the states $\catp$, $\catm$, $\ket{\mathcal{C}_{\alpha}^{-i}}$, and $\ket{\mathcal{C}_{\alpha}^{+i}}$ followed by a variable wait time $\Delta t$. The second set is used to symmetrize the readout contrast. \textbf{b}, \textbf{d} Decay curves of the indicated states. Open blue circles are experimental data, solid black lines are exponential fits with decay times given in the text. \textbf{e}, Pulse sequence for measuring the coherent state lifetimes. A $Y(\pm \pi/2)$ rotation brings the Fock qubit along its X-axis and is followed by mapping onto the cat qubit and a variable wait time $\Delta t$. After mapping back, another $\pi/2$ rotation is applied with a phase $\phi(\Delta t)=2\pi f\sub{det}\Delta t$, where $f\sub{det}$ is an effective detuning. \textbf{f}, Loss of coherence and leakage to higher excited states as a function of time for initialization in state $\xp$. Blue dots are the evolution of the resulting $\ket{0}$-state population of the Fock qubit using an effective detuning of $\SI{50}{\kilo \hertz}$ (dark blue) and $\SI{200}{\kilo \hertz}$ (light blue). The time $T_{e^{-1}}\approx\SI{110}{\micro \second}$ after which the contrast has decreased by a factor $1/e$ is indicated. Red dots show the population of Fock states $\ket{n>1}$ and are offset for clarity.  \textbf{g}, \textbf{h}, Time-dependence of the population fraction for the Fock states $\ket{n>1}$ using the pulse sequences shown in \textbf{a} (for $\catp$) and \textbf{b} (for $\ket{\mathcal{C}_{\alpha}^{-i}}$). Open blue circles are experimental data, solid black lines are exponential fits with rise times given in the text.}
\end{figure}

We measure the coherence time of the states along the Z- and Y-axis of the cat qubit using dispersive readout as shown in Fig.~\ref{SIfig:coh_disp}a, c: First, we initialize the Kerr-cat qubit in the state $\catp$ by ramping on the squeezing drive. Then we apply one of four operations (X(0)=$\mathds{1}$, $X(\pi)$, $X(\pi/2)$, and $X(-\pi/2)$) leaving the cat qubit in the state $\catp$, $\catm$, $\ket{\mathcal{C}_{\alpha}^{-i}}$, and $\ket{\mathcal{C}_{\alpha}^{+i}}$ respectively. After a variable wait time $\Delta t$, we perform another set of operations to reorient the cat qubit along its Z-axis, map back onto the Fock qubit and read out dispersively. In order to symmetrize the readout contrast, we apply the return-operations $X(0)$ as well as $X(\pi)$ for each of the first two states and $X(\pi/2)$ as well as $X(-\pi/2)$ for each of the second two states and take the difference between the thresholded measurement results. In all four obtained datasets (see Fig.~\ref{SIfig:coh_disp}b and Fig.~\ref{SIfig:coh_disp}d) we observe an exponential decay with respective decay times: ${\SI{2.62}{\micro\second}\pm\SI{0.06}{\micro\second}\xspace}$, ${\SI{2.56}{\micro\second}\pm\SI{0.05}{\micro\second}\xspace}$, ${\SI{2.53}{\micro\second}\pm\SI{0.05}{\micro\second}\xspace}$, and ${\SI{2.61}{\micro\second}\pm\SI{0.05}{\micro\second}\xspace}$ corresponding well to the decay times observed when using the cat-quadrature readout.

We measure the lifetime of the coherent states $\xpm$ by first initializing the Fock qubit along the X-axis with a $Y(\pm \pi/2)$ operation, then mapping onto the Kerr-cat qubit, waiting for a variable time $\Delta t$, mapping back onto the Fock qubit and finally performing a second $Y(\pm \pi/2)$ operation followed by dispersive readout. The second pulse is performed with a software detuning leading to the oscillations visible in the datasets shown in Fig.~\ref{SIfig:coh_disp}f. We show only the oscillating population of the $\ket{0}$ state for $\xp$ to avoid redundancy. The two sets of blue dots in the figure correspond to different detunings. Note that here we do not symmetrize the measurement and the curve decays to a value other than $0.5$, as expected because of leakage to higher excited states. Therefore, we do not fit it with an exponential decay. We can however mark the time at which its contrast has been reduced by a factor $e^{-1}$. This happens at $\mathrm{T}_{e^{-1}}\approx\SI{110}{\micro \second}$ corresponding well to the timescale found with cat-quadrature readout.

The dispersive measurement used here also gives us access to the populations of the higher excited states of the Fock qubit at the end of the experiment which reflect the leakage out of the cat-encoding space (see section~\ref{sec:gap}). We plot the population fraction of the states $\ket{n>1}$ as red dots in the figure. Small oscillations of this curve indicate a small error in the thresholding regions used in our single-shot measurement, but do not change the overall behavior. The curve indicates an increase and saturation of the population of higher excited states with a characteristic time $\mathrm{T_{>1}}\approx \SI{21}{\micro \second} \pm\SI{2}{\micro \second}$. The same quantities are plotted for the coherence measurements of $\catp$ and $\ket{\mathcal{C}_{\alpha}^{-i}}$ in Fig.~\ref{SIfig:coh_disp}g and Fig.~\ref{SIfig:coh_disp}h. The curves for $\catm$ and $\ket{\mathcal{C}_{\alpha}^{+i}}$ are very similar and are not shown. The constant offset in population is due to a small threshold error attributing some $\ket{0}$ and $\ket{1}$ state population to the $\ket{>1}$ states and does not affect the overall behavior of the curve as a function of $\Delta t$. An exponential fit to the data for all four states ($\catp$, $\catm$, $\ket{\mathcal{C}_{\alpha}^{-i}}$, and $\ket{\mathcal{C}_{\alpha}^{+i}}$) gives the respective rise time constants $\SI{14.6}{\micro \second}\pm\SI{0.8}{\micro \second}$, $\SI{14.1}{\micro \second}\pm\SI{0.9}{\micro \second}$, $\SI{15.4}{\micro \second}\pm\SI{1.1}{\micro \second}$, and $\SI{14.1}{\micro \second}\pm\SI{0.8}{\micro \second}$.

\newpage
\section{Decoherence processes and leakage to higher excited states}
\label{sec:decoherence}
In this section we will discuss the impact of different decoherence mechanisms on the coherence properties of the presented Kerr-cat qubit. We will first develop analytic expressions in the limit $p \to 1$ and for $\alpha \in \mathds{R}$. The action of the annihilation and creation operators on the cat qubit basis states is well approximated by (see section~\ref{sec:bloch})
\begin{align}
\hat{a}\catpm  &= \alpha \catmp\\
\hat{a}^{\dag}\catpm  &= \alpha \catmp + \ket{\psi_e^{\mp}} \\
\hat{a}\ket{\mathcal{C}_{\alpha}^{\mp i}}  &= \alpha \ket{\mathcal{C}_{\alpha}^{\pm i}}\\
\hat{a}^{\dag}\ket{\mathcal{C}_{\alpha}^{\mp i}}  &= \alpha \ket{\mathcal{C}_{\alpha}^{\pm i}} + \ket{\psi_e^{\pm i}} \\
\hat{a}\xpm  &= \pm \alpha \xpm\\
\hat{a}^{\dag} \xpm  &= \pm \alpha \xpm + \mathrm{D}(\pm \alpha)\ket{1}
\label{eq:jump}
\end{align}
where the states $\ket{\psi_e^{\mp}}=(\mathrm{D}(\alpha)\pm\mathrm{D}(-\alpha))\ket{1}$ correspond to the next exited states of the Hamiltonian~({\eqone}) introduced and spectroscopically measured in section~\ref{sec:gap}. In this section we have shown that a coherent drive can only induce transitions from the cat states to the excited states at their difference frequency $\omega_a-\omega\sub{gap}$ in the laboratory frame, with $\hbar \omega\sub{gap}=E\sub{gap}$). Similarly, only $\hat{a}^{\dag}$-noise with a spectral density around this frequency can cause excitation events. These are usually strongly suppressed as evidenced by the low thermal photon number of the undriven mode. The evolution of the system is described by the master equation
\begin{equation}
\label{eq:rho_full}
\dot{\hat{\rho}} = -\frac{i}{\hbar}[\hat{H}\sub{cat},\hat{\rho}] +\kappa_a(1+n\sub{th})\mathcal{D}[\hat{a}]\hat{\rho} +\kappa_a n\sub{th} \mathcal{D}[\hat{a}^{\dag}]\hat{\rho} + \kappa_{\phi,\mathrm{eff}} \mathcal{D}[\hat{a}^{\dag}\hat{a}]\hat{\rho},
\end{equation}
where $\mathcal{D}[\mathcal{O}]\hat{\rho}=\mathcal{O}\hat{\rho}\mathcal{O}^{\dag}-\frac{1}{2}\mathcal{O}^{\dag}\mathcal{O}\hat{\rho} -\frac{1}{2}\hat{\rho}\mathcal{O}^{\dag}\mathcal{O}$, $\kappa_a=1/T_1$ is the single photon loss rate of the nonlinear resonator,  $n\sub{th}$ its equilibrium thermal occupation number, and $\kappa_{\phi,\mathrm{eff}}$ is an effective dephasing rate.

In this description we use the ideal Hamiltonian $H\sub{cat}$ given in equation~({\eqone}), because we aim to describe the experiments presented in Fig.~{\figfour} where we have accounted for spurious drive-dependent frequency detunings. The effective dephasing rate $ \kappa_{\phi,\mathrm{eff}}$ will be used to give a phenomenological description of the corresponding noise, which in our flux-biased system likely has a $1/f$ frequency dependence instead of the white noise spectral density assumed here. It is possible to explicitly study the impact of colored noise on this type of qubit~\cite{Puri2018a}, but this is outside of the scope of this work.

\subsection{Photon loss and gain: Bit flips}
We will for now neglect leakage to the excited states as well as dephasing noise (setting $ \kappa_{\phi,\mathrm{eff}}=0$) and focus on the effect of photon loss and photon gain within the cat-encoding. In this case both operators only lead to flips of the states on the Z-axis and Y-axis of the cat qubit, while leaving the coherent states on the X-axis invariant.


In the basis of the even and odd cat states with respective populations $\rho_{00}$ and $\rho_{11}$ equation~(\ref{eq:rho_full}) yields the differential equations
\begin{align*}
\dot{\rho}_{00}(t)  &= \bar{n}\kappa_a(1+2n\sub{th})\rho_{11}(t) - \bar{n}\kappa_a(1+2n\sub{th})\rho_{00}(t)\\
\dot{\rho}_{11}(t)  &= \bar{n}\kappa_a(1+2n\sub{th})\rho_{00}(t) - \bar{n}\kappa_a(1+2n\sub{th})\rho_{11}(t)
\end{align*}
with the solutions
\begin{align*}
\rho_{00}(t)  &= \frac{1}{2} + \frac{1}{2}(\rho_{00}(0)-\rho_{11}(0))e^{-2\bar{n}\kappa_a(1+2 n\sub{th})t}\\
\rho_{11}(t)  &= \frac{1}{2} - \frac{1}{2}(\rho_{00}(0)-\rho_{11}(0))e^{-2\bar{n}\kappa_a(1+2 n\sub{th})t}
\end{align*}

These expressions show the action of the bit flip channel where any initial state decays to an equal statistical mixture of even and odd cat states with a time constant given by $2\bar{n}\kappa_a(1+2 n\sub{th})$, where $\bar{n}=\alpha^{2}$ is the average photon number in the cat state. The same result is found for the flips between the states $\ket{\mathcal{C}_{\alpha}^{\mp i}}$. Using the measured thermal photon number $n\sub{th}=\nth$ we would expect a bit-flip time of $\approx\SI{2.8}{\micro \second}$ which is slightly larger than the experimentally measured values. Heating to a value of $n\sub{th}=0.08$ (yielding $\approx\SI{2.6}{\micro \second}$) due to the application of the drives~\cite{Sank2016a} could account for this difference.

\subsection{Photon gain: Leakage to higher excited states and phase flips}

The definitions given in equations~(\ref{eq:jump}-\ref{eq:rho_full}) indicate that the process $\kappa_a n\sub{th} \mathcal{D}[\hat{a}^{\dag}]$, which causes leakage from the cat states to the excited states, happens at a rate $\kappa_a n\sub{th}$. We now focus on the timescale over which the corresponding populations equilibrate (as measured in Fig.~\ref{SIfig:coh_disp}).  In the approximation $n\sub{th} \ll 1$, the rate equations for the population $p_{c}$ of the cat states and the population $p_{e}$ of the first two excited states ($\ket{\psi_e^{\mp}}$) are given by:
\begin{align*}
\dot{p}_{e}(t)  &= \kappa_a n\sub{th}p_{c}(t) - \kappa_a(1+n\sub{th})p_{e}(t)\\
\dot{p}_{c}(t)  &= -\kappa_a n\sub{th}p_{c}(t) + \kappa_a(1+n\sub{th})p_{e}(t)
\end{align*}
which we solve with the initial conditions $p_{c}(0)=1$ and $p_{e}(0)=0$. This results in a time-dependence of the populations given by
\begin{align*}
p_{e}(t)  &= \frac{n\sub{th}}{1+2n\sub{th}}(1-e^{-\kappa_a(1+2 n\sub{th})t}) \\
p_{c}(t)  &= \frac{1}{1+2n\sub{th}} + \frac{n\sub{th}}{1+2n\sub{th}}(1+e^{-\kappa_a(1+2 n\sub{th})t})
\end{align*}
indicating that they reach equilibrium on a timescale of $1/(\kappa_a(1+2 n\sub{th}))\approx\SI{14}{\micro \second}$. This gives the correct order of magnitude, while underestimating the timescale found in experiment.

Finally, we note that for the photon number $\bar{n}=\nbarRest$ demonstrated in this work the excited states are above the energy barrier shown in Fig.~\ref{SIfig:gap}. This means that the excited states do not benefit from the same suppression of phase flips as the cat states. We therefore expect that thermal jumps to these excited states contribute significantly to the observed phase-flip rate.

\subsection{Dephasing: Leakage to higher excited states and phase flips}

We have so far neglected dephasing noise described by $\kappa_{\phi,\mathrm{eff}} \mathcal{D}[\hat{a}^{\dag}\hat{a}]$. In this section, we will discuss its influence qualitatively, while in the next section we will investigate it further through numerical simulations.

The effect of this type of noise within the cat Bloch sphere can be understood by adding a term $\Delta \hat{a}^\dagger \hat{a}$ to the Kerr-cat Hamiltonian~(\eqone), where $\Delta$ is some unknown fluctuating quantity with a spectral density of fluctuations given by the environmental noise. We analyze the effect of this noisy Hamiltonian term in two parts: rotations within the cat Bloch sphere, and leakage to higher excited states. Firstly, projected in the cat-qubit basis, this Hamiltonian reduces to $\Omega_z \hat{\sigma}_z$ where the Rabi rate around the Z-axis is $\Omega_z = -4\Delta|\alpha|^2 e^{-2|\alpha|^2}$ \cite{Puri2017}. This shows that the susceptibility of the cat qubit to \textit{any} frequency of $\hat{a}^\dagger \hat{a}$ noise is exponentially suppressed. Furthermore, we do not expect this noise to contribute to the bit flip rate, because the underlying operator conserves parity.

Additionally, the operator $\hat{a}^{\dag}\hat{a}$ can also cause leakage to the excited states, which leads to dephasing of the cat qubit for our experimental parameters as discussed in the previous section. It can be shown~\cite{Puri2018a}, that such leakage only occurs for noise with a spectral density around the gap frequency $\omega\sub{gap}$. This can be understood in a similar manner to the effect of a spin-locking experiment where a driven qubit becomes sensitive to noise only around the induced Rabi frequency \cite{yan2013}.

In our experiment we observe a reduction of the overall susceptibility to dephasing noise of the Kerr-cat qubit with respect to the Fock qubit as evidenced by the increase of the respective transverse decoherence times along the qubit X-axis.

\subsection{Numerical simulation}

In this section we simulate the evolution of the states on the six cardinal points of the cat-qubit Bloch sphere according to the master equation~(\ref{eq:rho_full}) and compare it to the data presented in Fig.~{\figfour} of the main text. We use the independently determined values of $K$ and $\epsilon_2$, but vary both $n\sub{th}$ and $\kappa_{\phi,\mathrm{eff}}$ to investigate the effect of these parameters on the predicted coherence properties.

We start out with $\kappa_{\phi,\mathrm{eff}}=0$ meaning that we neglect the suppressed impact of the dephasing noise as well as potential leakage due to this noise process. We observe that for a value of $n\sub{th}=\nth$, corresponding to the measured value in the undriven system, our simulation overestimates both the bit-flip and phase-flip times. An increase to a value of $n\sub{th}=0.08$ yields the correct bit-flip time as expected from our prior analysis, but does not reproduce the phase-flip time.

\begin{figure}[h]
	\includegraphics[angle = 0, width = \figwidthWide]{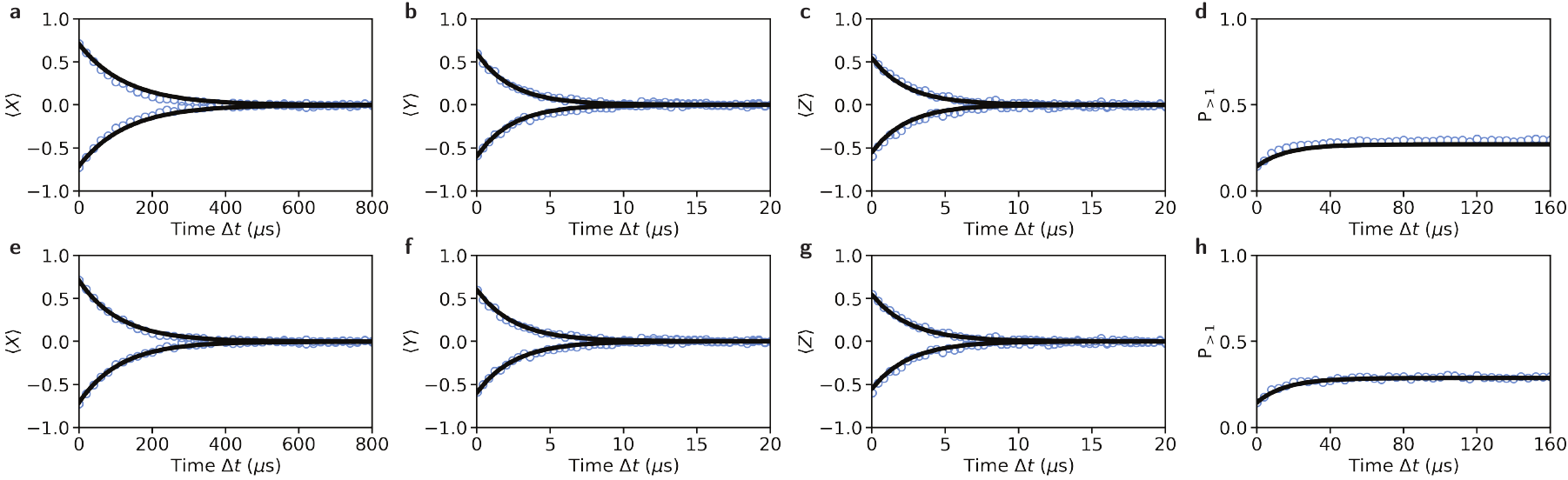}
	\caption{\label{SIfig:decoh_sim}  \textbf{Comparison of simulated decoherence to data. a},\textbf{b},\textbf{c}, Decay curves of the cat-qubit states $\ket{\pm X}$, $\ket{\pm Y}$, and $\ket{\pm Z}$. Open blue circles are the data presented in Fig.~{\figfour}. Solid black lines are the corresponding expectation values of the cat-qubit state found from simulation of equation~(\ref{eq:rho_full}) for $n\sub{th}=0.12$ and $\kappa_{\phi,\mathrm{eff}}/2\pi=0$ and scaled to match the experimental measurement contrast. \textbf{d}, Increase of the excited state population. Open blue circles are data shown in  Fig.~\ref{SIfig:coh_disp}. The solid black line is a simulation $n\sub{th}=0.12$ and $\kappa_{\phi,\mathrm{eff}}/2\pi=0$, offset to match the value of the data at $\Delta t=0$, but not scaled. \textbf{e},\textbf{f},\textbf{g}, Same experimental data as in \textbf{a},\textbf{b},\textbf{c}. Simulation results were obtained for $n\sub{th}=0.08$ and $\kappa_{\phi,\mathrm{eff}}/2 \pi=\SI{230}{\hertz}$ and scaled to match the measurement contrast. \textbf{h}, Same experimental data as in \textbf{d}. Simulation data is shown for $n\sub{th}=0.08$ and $\kappa_{\phi,\mathrm{eff}}/2 \pi=\SI{230}{\hertz}$, offset but not scaled.}   
\end{figure}

A reasonably good agreement can only be found for a value of $n\sub{th}=0.12$ as shown in Fig.~\ref{SIfig:decoh_sim}. In panels \textbf{a}, \textbf{b}, and \textbf{c} of the figure we compare the experimental data to the simulation results. To this end, we scale the latter to the readout contrast of the former. The simulation finds a bit-flip time of $\approx\SI{2.4}{\micro \second}$ and a phase-flip time of $\approx\SI{130}{\micro \second}$, thus very slightly underestimating the former while overestimating the latter. In Fig.~\ref{SIfig:decoh_sim}d we compare the leakage to the excited states found in simulation by evaluating ${1-\mathrm{tr}(\ket{\mathcal{C}_{\alpha}^{+}}\bra{\mathcal{C}_{\alpha}^{+}} \hat{\rho}) - \mathrm{tr}(\ket{\mathcal{C}_{\alpha}^{-}}\bra{\mathcal{C}_{\alpha}^{-}} \hat{\rho})}$ to the average of the measured $P\sub{>1}$-datasets presented in section~\ref{sec:disp}. As described in this section, we attribute the offset in the data to an imprecision in the thresholding of the single-shot measurement that does not change the rise-time or overall scaling of the population increase. In order to compare simulation to experiment we therefore offset the simulation result such that it matches the data at time $\Delta t=0$, but do not rescale it. The leakage rise time is found to be $\approx \SI{16}{\micro \second}$. The population increase is slightly smaller than found in the experiment.

In a second simulation we choose the value $n\sub{th}=0.08$ which we expect for the measured bit-flip time and add a small amount of effective dephasing noise with $\kappa_{\phi,\mathrm{eff}}/2\pi=\SI{230}{\hertz}$. The corresponding simulation results reproduce the data well with a bit-flip time of $\approx\SI{2.6}{\micro \second}$, a phase-flip time of $\approx\SI{110}{\micro \second}$ and a leakage rise time of $\approx{16}{\micro \second}$. They are presented in Fig.~\ref{SIfig:decoh_sim}e,f,g,h.

While these results are only based on a phenomenological description of the system (in particular with regard to the fact that the spectral-density of the dephasing noise in our experiment is most likely not uniform) they do give an intuition on the limiting factors in the current experiment. It seems probable that a large part of the remaining phase flips comes from a combination of leakage to states outside of the cat encoding and the fact that these states are above the energy barrier. Strategies to reduce the impact of this effect include increasing the ratio between $g_3$ and $K$ in order to reduce effective pump strengths, as well as introducing two-photon dissipation~\cite{Leghtas2015, Mirrahimi2014, Puri2018a} to counteract the heating. An increase in photon number by a factor of two would bring the first excited states below the energy barrier~\cite{Puri2018a} and, combined with the previously mentioned strategies, would make the system insensitive to leakage to these levels. A more in-depth study of these effects as a function of $\alpha$ will be the topic of future work.

\end{document}